\documentclass[useAMS,usenatbib]{mnras}
\usepackage{graphicx,natbib,amssymb,amsmath,stmaryrd,makecell}
\usepackage{soul}
\usepackage{dblfloatfix}
\usepackage{subcaption}
\usepackage[T1]{fontenc}
\usepackage{ae,aecompl}
\usepackage{fixltx2e}
\usepackage[english]{babel}
\usepackage{graphicx,amsmath}
\usepackage{epstopdf}
\usepackage{epsf,color}
\usepackage[mathscr]{eucal}
\usepackage{amsmath}
\usepackage{amssymb,amsfonts}
\usepackage{natbib}
\usepackage{graphicx}
\usepackage{txfonts}
\usepackage{dsfont}
\definecolor{Mygreen}{rgb}{0.00, 0.5, 0.5}
\definecolor{Mypink}{rgb}{1.0, 0.0, 0.5}
\definecolor{Myblue}{rgb}{0.00, 0.2, 0.8}
\definecolor{Myred}{rgb}{0.80, 0.2, 0.0}
\usepackage{float}
\usepackage{color}
\def\simlt{\lower.5ex\hbox{$\; \buildrel < \over \sim \;$}}
\def\simgt{\lower.5ex\hbox{$\; \buildrel > \over \sim \;$}}

\bibpunct{(}{)}{ and}{a}{}{,}
\newfont{\gwpfont}{cmssq8 scaled 1000}

\title[Impact of the mean pressure profile on tSZ cosmology]{Impact of the mean pressure profile of galaxy clusters on the cosmological constraints from the $Planck$ tSZ power spectrum}

\pubyear{2019}

\author[F.~Ruppin \emph{et al.}]{
F.~Ruppin,$^{1,2}$\thanks{E-mail: ruppin@mit.edu}
F.~Mayet,$^{1}$
J.F.~Mac\'ias-P\'erez$^{1}$
and L.~Perotto$^{1}$
\\
$^{1}$Univ. Grenoble Alpes, CNRS, LPSC/IN2P3, 53 avenue des Martyrs, 38000 Grenoble, France\\
$^{2}$Kavli Institute for Astrophysics and Space Research, Massachusetts Institute of Technology, Cambridge, MA 02139, USA
}

\begin{document}

\def\aj{AJ}%
\def\araa{ARA\&A}%
\def\apj{ApJ}%
\def\apjl{ApJ}%
\def\apjs{ApJS}%
\def\aap{Astron. Astrophys.}%
 \def\aapr{A\&A~Rev.}%
\def\aaps{A\&AS}%
\def\mnras{MNRAS}
\def\ssr{SSRv}
\def\nat{Nature}
\def\jcap{JCAP}

\def\Mgv{M_{\rm g,500}}
\def\Mg{M_{\rm g}}
\def\YX {Y_{\rm X}}
\def\LXv {L_{\rm X,500}}
\def\TX {T_{\rm X}}
\def\fgv {f_{\rm g,500}}
\def\fg  {f_{\rm g}}
\def\kT {{\rm k}T}
\def\ne {n_{\rm e}}
\def\Mv {M_{\rm 500}}
\def \Rv {R_{500}}
\def\keV {\rm keV}
\def\Yv{Y_{500}}

\def\MT {$M$--$T_{\rm X}$}
\def\MYX {$M$--$Y_{\rm X}$}
\def\MMg {$M_{500}$--$M_{\rm g,500}$}
\def\MgT {$M_{\rm g,500}$--$T_{\rm X}$}
\def\MgY {$M_{\rm g,500}$--$Y_{\rm X}$}

\def\msol {{\rm M_{\odot}}}

\def\lesssim{\mathrel{\hbox{\rlap{\hbox{\lower4pt\hbox{$\sim$}}}\hbox{$<$}}}}
\def\gtrsim{\mathrel{\hbox{\rlap{\hbox{\lower4pt\hbox{$\sim$}}}\hbox{$>$}}}}

\def\psz{PSZ2\,G144.83$+$25.11}

% satellites
\def\xmm{XMM-{\it Newton}}
\def\planck{{\it Planck}} 
\def\chandra{{\it Chandra}}
\def \rosat {\hbox{\it ROSAT}}
\newcommand{\excpres}{{\gwpfont EXCPRES}}
\newcommand{\ma}[1]{\textcolor{red}{{ #1}}}
%###############################################################################################
%##########################              START THE PAPER             ##########################################
%###############################################################################################
\label{firstpage}
\pagerange{\pageref{firstpage}--\pageref{lastpage}}
\maketitle

\begin{abstract}
Cosmological analyses based on surveys of galaxy clusters observed through the Sunyaev-Zel'dovich (SZ) effect strongly rely on the mean pressure profile of the cluster population. A tension is currently observed between the cosmological constraints obtained from the analyses of the CMB primary anisotropies and those from cluster abundance in SZ surveys. This discrepancy may be explained by a wrong estimate of the hydrostatic bias parameter that links the hydrostatic mass to the true mass of galaxy clusters. However, a variation of both the amplitude and the shape of the mean pressure profile could also explain part of this tension. We analyze the effects of a modification of this profile on the constraints of the $\sigma_8$ and $\Omega_m$ parameters through the analysis of the SZ power spectrum measured by the \planck\ collaboration. We choose two mean pressure profiles that are respectively lower and higher than the one obtained from the observation of nearby clusters by \planck. The selection of the parameters of these two profiles is based on the current estimates of the pressure and gas mass fraction profile distributions at low redshift. The cosmological parameters found for these two profiles are significantly different from the ones obtained with the \planck\ pressure profile. We conclude that a ${\sim}15\%$ decrease of the amplitude of the mean normalized pressure profile would alleviate the tension observed between the constraints of $\sigma_8$ and $\Omega_m$ from the CMB and cluster analyses without requiring extreme values of the mass bias parameter.
\end{abstract}


%#############################################################################################
%##########################                             INTRODUCTION                               ##########################%#############################################################################################
\section{Introduction}\label{sec:Introduction}
%---------- Cosmology with galaxy cluster
The abundance of galaxy clusters as a function of mass and redshift is a very powerful cosmological probe. Indeed, it is both sensitive to the primordial matter power spectrum and to the growth of structures across the whole history of the universe \citep[\emph{e.g.}][]{voi05}. The cosmological constraints obtained from the analysis of the cluster abundance are complementary to those estimated from other cosmological probes such as the primary anisotropies of the cosmic microwave background \citep[CMB;][]{pla18}, type Ia supernovae \citep{per97} or baryon acoustic oscillations \citep[BAO;][]{and14} for two main reasons. On the one hand, the degeneracies between the cosmological parameters estimated from the study of galaxy clusters are nearly orthogonal to those of the other probes. Multi-probe analyses can therefore lead to very tight cosmological constraints \citep{man15}. On the other hand, galaxy clusters enable studying the most recent state of the matter density field. Therefore, the comparison between the cosmological constraints derived from high-redshift probes such as the CMB with the ones obtained through the analysis of cluster surveys allows us to test our current model of structure formation \citep[\emph{e.g.}][]{mis18}.\\
%---------- The tSZ effect to do cosmology
The tSZ effect has been shown to be an excellent observable in order to establish nearly mass-limited cluster samples up to very high redshift \citep[\emph{e.g.}][]{boc18}. The integrated tSZ flux $Y$ measured for each cluster detected in a millimeter survey is proportional to the thermal energy content of the intracluster medium (ICM). Despite its lack of accuracy, due to the presence of non-thermal pressure support within the ICM, this observable provides a high precision proxy for the mass $M$ of galaxy clusters \citep[\emph{e.g.}][]{pla14}. However, accurate measurements of the integrated tSZ flux require either high exposure X-ray observations in order to deproject both the density and the temperature profiles up to high redshift or high angular resolution tSZ observations from the core up to the virial radius of galaxy clusters \citep[see][for a detailed review on ICM physics with spatially resolved SZ observations]{mro19}. Such observations have been and are currently realized on limited cluster samples in order to provide both the $Y{-}M$ scaling relation and the mean normalized pressure profile that are essential for tSZ cosmological analyses \citep[see \emph{e.g.}][]{arn10,pla13,per18}.\\
%---------- Current constraints with clusters
Several catalogs of galaxy clusters have been established from the tSZ observations realized by the \planck\ satellite \citep{pla16}, the South Pole Telescope \citep[SPT;][]{ble15}, and the Atacama Cosmology Telescope \citep[ACT;][]{has13} surveys. The analysis of the cluster abundance based on these catalogs or the study of the tSZ angular power spectrum \citep{pla16b} have enabled estimating cosmological parameters such as the amplitude of the linear matter power spectrum at a scale of $8h^{-1}$Mpc, $\sigma_8$, and the total matter density of the universe $\Omega_m$. However, these analyses lead to cosmological constraints that are in tension with the ones obtained from the analysis of the power spectrum of the CMB temperature anisotropies \citep{pla16c}.\\
%---------- Current limitations
We assume in this study that this tension is not due to a limit in the standard $\Lambda$CDM model but is caused by a biased estimation of the thermodynamic properties of galaxy clusters\footnote{We do not consider the impact of potential selection biases in this paper.}. The observed disagreement may be due to a combination of three different systematic effects. Two of them are associated with the way we link the tSZ observable to the cluster mass and the third one is related to the measurement of the tSZ observable itself. The most studied one corresponds to a wrong estimate of the hydrostatic bias parameter $b$. This paramerter links the value of the mass of galaxy clusters obtained from X-ray and tSZ observations under the assumption of hydrostatic equilibrium to their true mass in the $Y{-}M$ scaling relation \citep[see the detailed review by][for more information on systematic effects on cluster mass estimates]{pra19}. Many analyses have already been carried out to show that an incorrect estimation of the hydrostatic bias parameter could result in the observed cosmological tension \citep[\emph{e.g.}][]{pla16c,pla16b}. As shown in \cite{sal19}, a value of $b = 0.38 \pm 0.05$ is required to cancel the discrepancy between the constraints of $\sigma_8$ and $\Omega_m$ obtained by the analysis of CMB primary anisotropies and cluster abundance. However, the current estimates of the hydrostatic bias measured with different cluster samples seem to favour a much lower value of the hydrostatic bias parameter. Its average value presented in \cite{sal18} is given by $b = 0.2 \pm 0.08$. A second source of systematic effect in the relation between the tSZ observable and the cluster mass may arise from a mass and redshift dependence of the slope of the $Y{-}M$ scaling relation. Finally, a significant part of the observed discrepancy between the CMB and cluster cosmological parameters may also be due to an evolution of the mean normalized pressure profile of galaxy clusters with mass and redshift. In this case, the tSZ observable measured by assuming that all galaxy clusters share the same normalized pressure profile may also be biased on average.\\
In this paper, we choose to focus our analysis on this last systematic effect. Thus, our goal is to describe how a modification of the mean normalized pressure profile of galaxy clusters with respect to the ones that are currently used in tSZ cosmological analyses can affect the constraints on $\sigma_8$ and $\Omega_m$.\\
%---------- Importance of the pressure profile
In cluster count analyses, the mean normalized pressure profile is used in order to measure the integrated tSZ flux of each cluster that has been detected in the survey \citep[\emph{e.g.}][]{mel06}. In analyses based on the tSZ angular power spectrum, the shape of the mean normalized pressure profile has a significant impact at high multipole, and its amplitude is directly proportional to the one of the power spectrum model used in order to fit the data \citep[\emph{e.g.}][]{bol18}. Therefore a wrong estimate of both the shape and the amplitude of the mean normalized pressure profile of the cluster population may lead to a biased estimation of cosmological parameters from the analyses of the cluster abundance in tSZ surveys. The most commonly used mean normalized pressure profiles in tSZ cosmological analyses have been estimated using cluster samples at high mass and low redshift \citep{arn10,pla13}. However, the true mean normalized pressure profile of the cluster population could be significantly different if deviations from the self-similar hypothesis are observed in different regions of the mass-redshift plane. Differences in the mean normalized pressure profile in distinct redshift intervals have already been observed by \cite{mcd14} for example. Furthermore, \cite{ram15} have shown that the tSZ angular power spectrum measured by SPT \citep{geo15} seems to favor a redshift evolution of the pressure profile parameters if we assume the cosmological parameters from \cite{kom11} to be accurately measured. In addition, the cluster samples used to calibrate both the mean normalized pressure profile and the scaling relation may not be representative of the cluster samples used in cosmological analyses because of selection biases.\\
\begin{figure*}
\centering
\includegraphics[height=7cm]{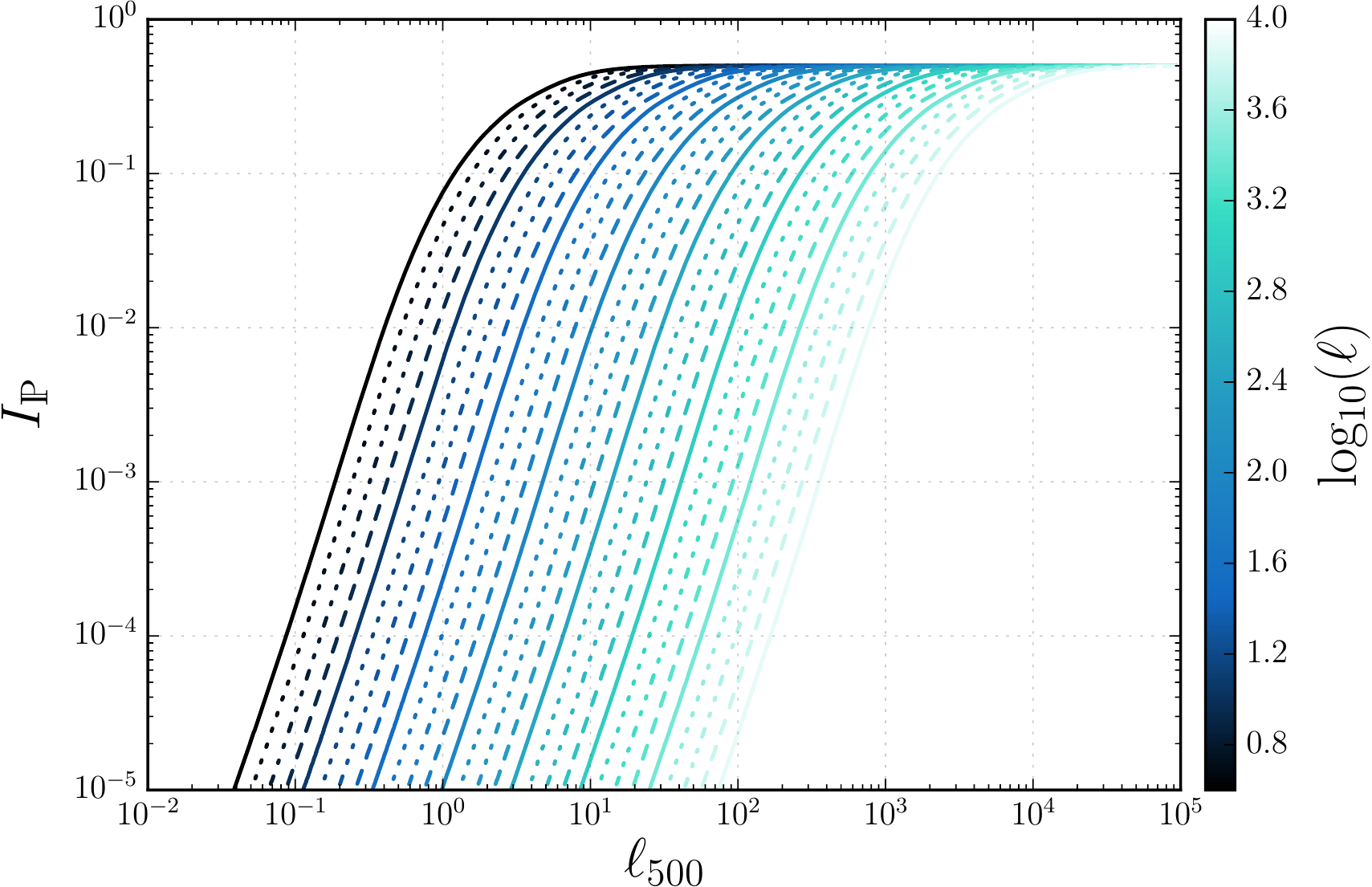}
\caption{Tabulated values of the $I_{\mathds{P}}$ function given by Eq. (\ref{eq:tabulated_y}) by considering the mean normalized pressure profile measured by the \planck\ collaboration \citep{pla13} as a function of $\ell_{500}$. The function has been computed for different multipoles $\ell$ given by the colorbar.}
\label{fig:tabulated_profile}
\end{figure*}One of the goals of the NIKA2 \citep{ada18} tSZ large program is to characterize the potential redshift evolution of the mean normalized pressure profile \citep{com16,per18}. The analysis of the first cluster observed by NIKA2, PSZ2\,G144.83$+$25.11 at $z=0.58$, has shown that cluster substructures can have a significant impact on the estimate of the ICM pressure profile \citep{rup18}. Furthermore, the prospective study of the NIKA2 tSZ large program based on the \emph{Marenostrum MUltidark SImulations of galaxy Clusters} \citep{sem12} has shown that the properties of the distribution of normalized pressure profiles may vary significantly with the fraction of morphologically disturbed clusters \citep{rup19}. Therefore, it is essential to characterize the impact of a potential modification of the mean normalized pressure profile on the constraints of cosmological parameters before we know exactly how to implement its potential mass and redshift evolution in cosmological analyses.\\
%---------- tSZ power spectrum analysis
In this paper, we estimate the implications of a modification of the mean normalized pressure profile of the cluster population on the constraints of the $\sigma_8$ and $\Omega_m$ parameters by analyzing the angular power spectrum of the tSZ effect measured by the \planck\ collaboration \citep{pla16b}. We define two extreme mean normalized pressure profiles that are respectively lower and higher than the one obtained from the observations of 62 nearby clusters by the \planck\ collaboration \citep{pla13}. For that purpose, we use the current knowledge on the distributions of normalized pressure profiles and gas mass fraction profiles observed at low redshift \citep[\emph{e.g.}][]{eck13,eck19}. Three Markov chain Monte Carlo (MCMC) analyses are performed in order to fit the \planck\ data with a tSZ angular power spectrum model based on these profiles. The cosmological constraints obtained from these analyses allow us to discuss the impact of a modification of the mean normalized pressure profile on the estimates of $\sigma_8$ and $\Omega_m$.\\
%---------- Paper organization
This paper is organized as follows. The different components of the tSZ power spectrum model and its dependence with cosmological parameters are described in Sect. \ref{sec:tsz_power_spec}. We present the data set that is considered in our analysis, \emph{i.e.} the \planck\ tSZ power spectrum measured from the whole sky $y$-map, in Sect. \ref{sec:planck_tsz_power}. We then describe the methodology that we have used in order to define the two mean normalized pressure profiles that enclose the one measured at low redshift by the \planck\ collaboration in Sect. \ref{sec:evol_mean_P}. The analysis of the \planck\ tSZ power spectrum based on these profiles is detailed in Sect. \ref{sec:MCMC_planck}. We discuss the implications of a modification of the mean normalized pressure profile on the estimation of cosmological parameters in this type of analysis in Sect. \ref{sec:results}. Finally, we present our conclusions in Sect. \ref{sec:Conclusions}. Throughout this study we assume a flat $\Lambda$CDM cosmology following the latest \planck\ results \citep{pla18}: $\Omega_b h^2 = 0.0224$, $n_s = 0.965$, $\tau = 0.054$, and $N_{\mathrm{eff}} = 2.99$. The $\sigma_8$, $\Omega_m$, and $H_0$ parameters are free in our analysis\footnote{The $\Omega_{\Lambda}$ parameter varies accordingly to keep a null curvature}.

\section{The tSZ angular power spectrum}\label{sec:tsz_power_spec}

\subsection{The thermal Sunyaev-Zel'dovich effect}\label{subsec:tsz_effect}

The thermal Sunyaev-Zel'dovich effect \citep{sun72,sun80} is caused by the inverse Compton scattering of CMB photons on energetic ICM electrons. It induces a variation of the apparent brightness of the CMB towards any line of sight that passes through a reservoir of hot plasma. This variation is given by:
\begin{equation}
        \frac{\Delta I_{tSZ}}{I_0} = y \, f(\nu, T_e),
\label{eq:deltaI}
\end{equation}
where the Compton parameter $y$ characterizes the amplitude of the spectral distortion, $f(\nu, T_e)$ is the frequency dependence of the tSZ spectrum \citep{bir99,car02}, and $T_e$ is the electronic temperature of the ICM. The Compton parameter in a given direction $\hat{n}$ is expressed as:
\begin{equation}
        y(\hat{n}) = \frac{\sigma_{\mathrm{T}}}{m_{e} c^2} \int P_{e} \, dl,
        \label{eq:y_compton}
\end{equation}
where $P_{e}$ is the ICM pressure distribution, $m_{e}$ is the electron mass, $c$ the speed of light, and $\sigma_{\mathrm{T}}$ the Thomson scattering cross section. The integrated Compton parameter $Y$ is given by the spherical integral of the ICM pressure distribution. It has been shown to be a low-scatter mass proxy for galaxy clusters \citep[e.g.][]{pla11}. In this paper, all the integrated quantities that characterize the ICM properties are obtained by considering the upper integration limit $\rm{R_{500}}$. This limit corresponds to the cluster radius for which the mean cluster density is $500$ times the critical density of the universe.

\subsection{Model of tSZ power spectrum}\label{subsec:model_tsz_power}

The tSZ power spectrum depends both on the geometric properties of the universe via the comoving volume element $d^2V/dzd\Omega$ and on the large scale structure formation processes via the mass function and the ICM pressure profile. We choose to model the pressure profile of galaxy clusters using a generalized Navarro-Frenk-White profile \citep[gNFW, ][]{nag07}:
\begin{equation}
P_e(x \, ; \, M_{500},z) = P_{500}(M_{500},z) \times \mathds{P}(x)
\label{eq:cosmo_P_prof}
\end{equation}
where $x=r/R_{500}$ and $\mathds{P}(x)$ is the normalized pressure profile\footnote{\emph{i.e.} it does not depend on the considered mass and redshift.} given by:
\begin{equation}
\mathds{P}(x) = \frac{P_0}{(c_{500}x)^{\gamma}[1+(c_{500}x)^{\alpha}]^{(\beta - \gamma)/\alpha}}
\label{eq:cosmo_normalized_P}
\end{equation}
The shape and amplitude of the normalized pressure profile are entirely defined by the parameters $P_0$, $c_{500}$, $\alpha$, $\beta$ and $\gamma$ in Eq. (\ref{eq:cosmo_normalized_P}). The characterization of the impact of the value of these parameters on the estimates of $\sigma_8$ and $\Omega_m$ obtained by the analysis of the power spectrum of the tSZ effect is the goal of the study developed in Sect. \ref{sec:results}. The scaling factor between the pressure content and the cluster total mass is given by \citep{arn10}:
\begin{equation}
P_{500} = 1.65 \times 10^{-3} \, E_z^{8/3} \, \left[ \frac{(1-b) \, M_{500}}{3 \times 10^{14} \, h_{70}^{-1}~\mathrm{M_{\odot}}}\right]^{2/3+0.12} h_{70}^2~\mathrm{keV} \, \mathrm{cm^{-3}}
\label{eq:P500}
\end{equation}
where $b$ is the hydrostatic bias parameter that links the true mass $M_{500}$ to the one given under the assumption of hydrostatic equilibrium, $E_z$ is the ratio of the Hubble constant at redshift $z$ to its present value $H_0$, and $h_{70} = H_0 / [70~\mathrm{km/s/Mpc}]$.
\begin{figure*}
\centering
\includegraphics[height=7cm]{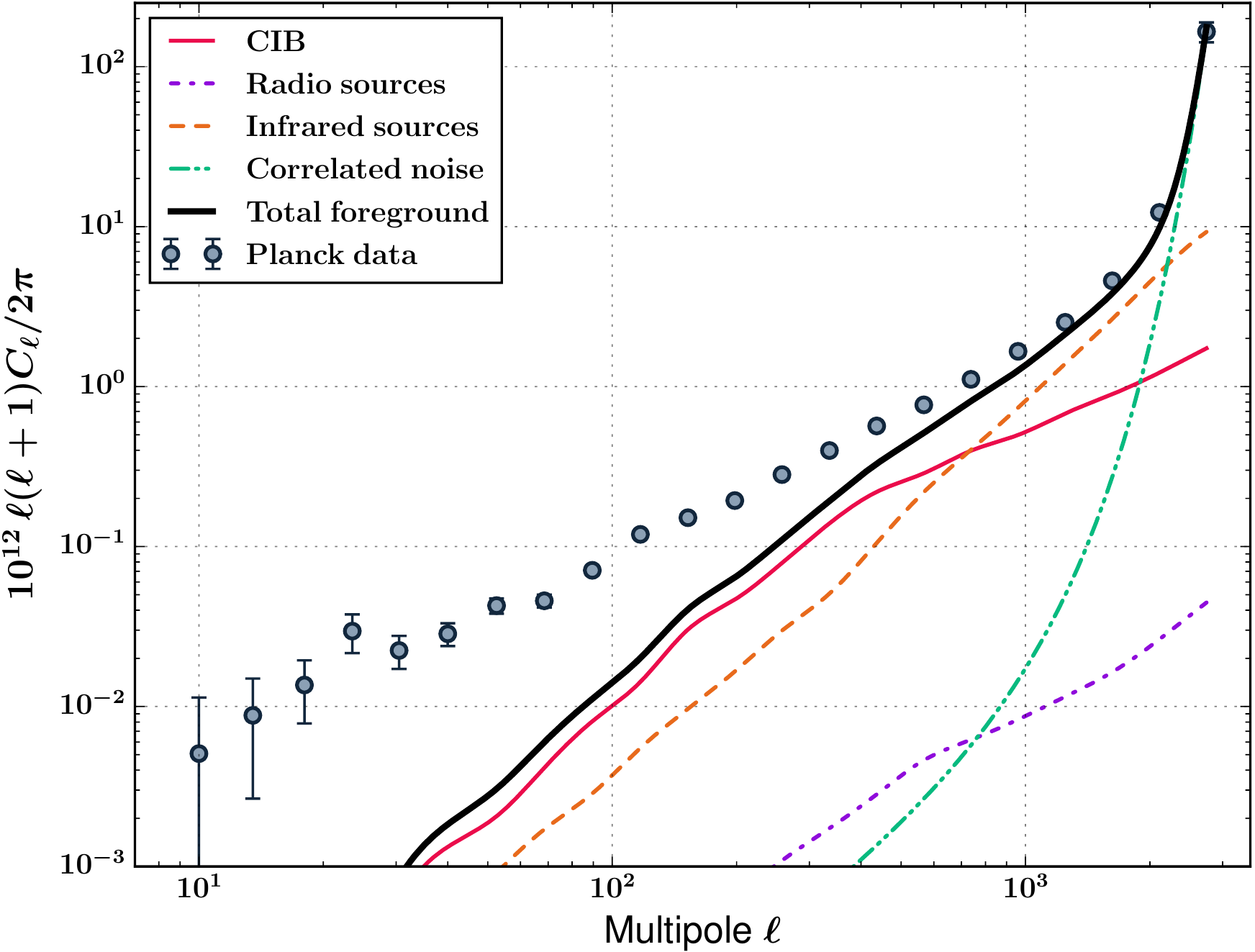}
\caption{\planck\ tSZ angular power spectrum measured from the full sky $y$-map \citep{pla16b} (grey points). The power spectra of the contaminants to the tSZ signal are given for the cosmic infrared background (red), radio sources (purple), infrared sources (orange), and spatially correlated residual noise (green). The sum of the contaminant power spectra is given by the black line.}
\label{fig:planck_power_spec}
\end{figure*}

As shown in \cite{kom99}, the contribution of the two-halo term in the angular power spectrum of the tSZ effect is negligible given the precision of the current tSZ measurements. The power spectrum of the tSZ effect is therefore modeled entirely by the one-halo component:
\begin{equation}
C_{\ell}^{tSZ} = \int \frac{d^2V}{dzd\Omega} \, dz \int \frac{dn}{dM_{500}} \, \left|\frac{4\pi R_{500}}{\ell_{500}^2}  \frac{\sigma_T}{m_ec^2} \, P_{500} \, I_{\mathds{P}}(\ell_{500})\right|^2 \, dM_{500} 
\label{eq:ClSZ_cosmo_analysis}
\end{equation}
where $dn/dM_{500}$ is the mass function that gives the expected halo number density for a given mass. In this paper, we choose to use the multiplicity function of \cite{tin08} because it corresponds to the function used by the \planck\ collaboration for cosmological analyses based on the study of the abundance of galaxy clusters \citep[\emph{e.g.}][]{pla16c}. The normalized two dimensional Fourier transform of the mean pressure profile $I_{\mathds{P}}(\ell_{500})$ is given under the Limber's approximation by \citep{kom02}:
\begin{equation}
I_{\mathds{P}} = \int x^2 \frac{\mathrm{sin}(\ell x/\ell_{500})}{\ell x/\ell_{500}} \mathds{P}(x) \, dx
\label{eq:tabulated_y}
\end{equation}
The values of this integral for a given cosmological model, for the considered multipoles $\ell$ and different values of $\ell_{500} \equiv D_A/R_{500}$, where $D_A$ is the angular diameter distance, are saved in a file at the beginning of each analysis (see Sect. \ref{sec:MCMC_planck}). The bounds of the $I_{\mathds{P}}$ integral correspond to a range of radii from $10^{-10}$ to $7R_{500}$ in order to integrate the entire pressure distribution from the core to the periphery of the halos. We represent in Fig. \ref{fig:tabulated_profile} the values of $I_{\mathds{P}}$ computed for 30 values of $\ell$ between 4 and $10^4$ as a function of $\ell_{500}$ for the mean pressure profile estimated by the \planck\ collaboration at low redshift \citep{pla13}. Knowing these values for a fixed normalized pressure profile simplifies the calculation of the tSZ power spectrum model given in Eq. (\ref{eq:ClSZ_cosmo_analysis}). Indeed, we just need to interpolate $I_{\mathds{P}}(\ell_{500})$ at the considered value of $\ell_{500}$ instead of integrating the pressure profile at each step of the analysis detailed in Sect. \ref{sec:MCMC_planck}.

\subsection{Cosmological parameters from the tSZ power spectrum}\label{subsec:cosmo_tsz_power}

In this paper, the comoving volume element as well as the mass function are both computed using the \texttt{hmf} python library for a given set of cosmological parameters \citep{mur13}. 

We carry out an analysis in order to study the variations of the amplitude and the shape of the power spectrum given by Eq. (\ref{eq:ClSZ_cosmo_analysis}) as a function of cosmological parameters \citep[see \emph{e.g.} Fig. 2 in][]{bol18}. The mean pressure profile constrained by the study of 62 low redshift clusters observed by \planck\ is considered in order to estimate $I_{\mathds{P}}$ \citep{pla13}. The $\Lambda$CDM fiducial model used in this study is the one constrained by the \planck\ collaboration \citep{pla18} (see Sect. \ref{sec:Introduction}). We focus our analysis on the values of the cosmological parameters $\sigma_8$, $\Omega_m$, $h = H_0 / [100~\mathrm{km/s/Mpc}]$, and the one of the hydrostatic bias parameter $b$. The impact of the value of these four parameters on the power spectrum of the tSZ effect is studied by applying Eq. (\ref{eq:ClSZ_cosmo_analysis}) for different variations of the fiducial cosmological model.\\
\begin{figure*}
\centering
\includegraphics[height=7cm]{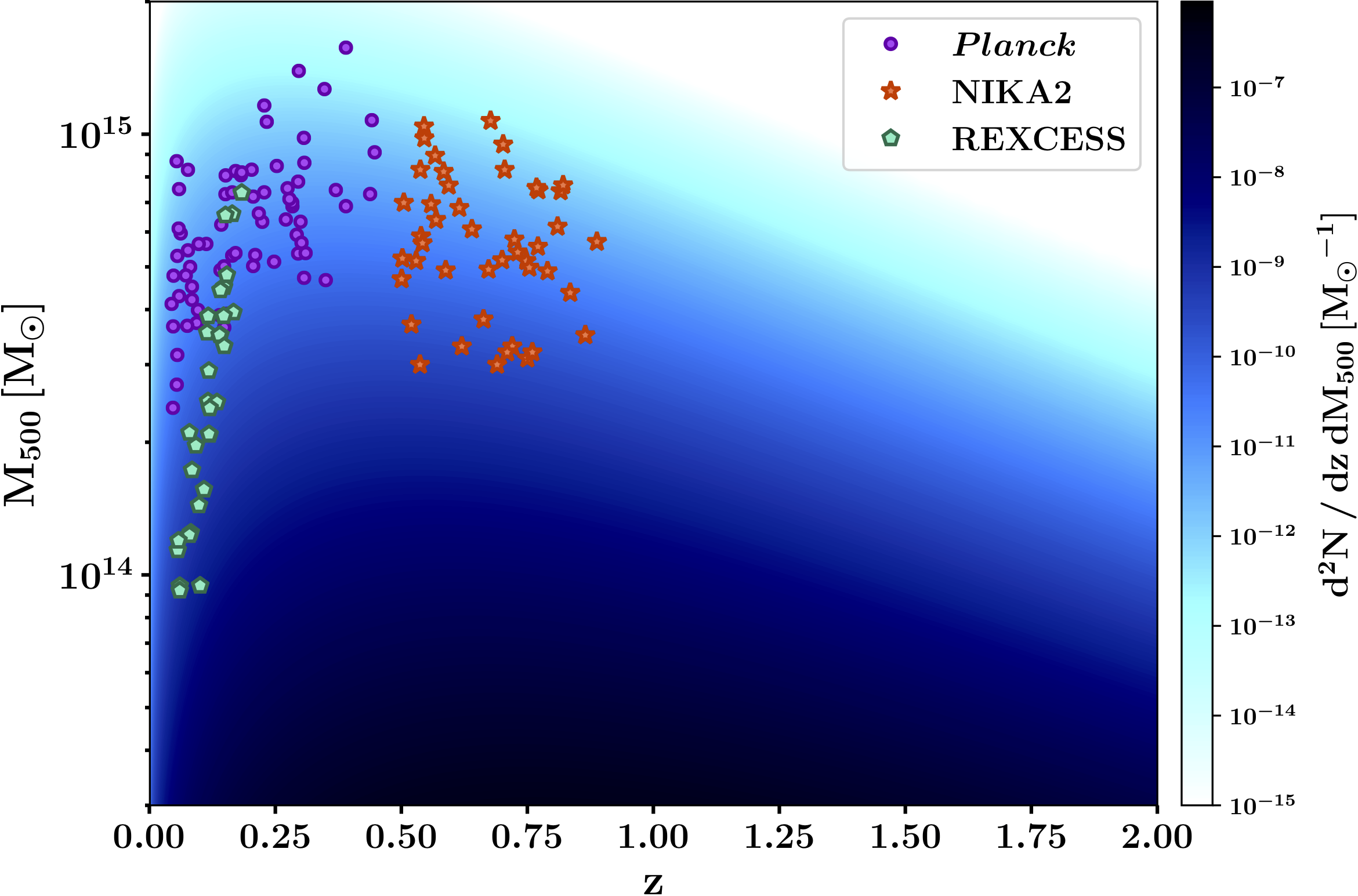}
\caption{Distribution of the 62 \planck\ clusters used in order to estimate the mean normalized pressure profile at low redshift (purple) in the mass-redshift plane. The NIKA2 and REXCESS samples are also shown in orange and green respectively. The different shades of blue give the expected cluster abundance, \emph{i.e.} the cluster number per unit of mass and redshift computed from Eq. (\ref{eq:expec_num}).}
\label{fig:context_cosmo}
\end{figure*}

We observe that the amplitude of the power spectrum of the tSZ effect increases with the value of $\sigma_8$. This is expected because the parameter $\sigma_8$ characterizes the normalization of the linear matter power spectrum. The amplitude of the power spectrum of the tSZ effect also increases with $\Omega_m$. We therefore expect an anti-correlation between the parameters $\sigma_8$ and $\Omega_m$ constrained by fitting the measured tSZ power spectrum (cf. Sect. \ref{sec:results}). The amplitude of the tSZ power spectrum in the multipole range considered in this study decreases as the value of the $h$ and $b$ parameters increases. This is expected because an increase of the value of $h$ implies that the expansion rate of the universe increases and therefore slows down the formation of the large-scale structures by gravitational collapse. In addition, as shown in Eq. (\ref{eq:P500}), an increase of the hydrostatic bias $b$ means that the normalization parameter $P_{500}$ is lower for a given cluster mass. We therefore expect a variation of the amplitude of the tSZ power spectrum that is proportional to $\sigma_8$ and $\Omega_m$ and inversely proportional to $h$ and $b$. In the rest of this study, we will use the following parameter in order to characterize the amplitude of the power spectrum of the tSZ effect:
\begin{equation}
F = \sigma_8 \, (\Omega_m/B)^{0.40} \, h^{-0.21}
\label{eq:defintion_F}
\end{equation}
This combined parameter, introduced by Bolliet \emph{et al. } \citep{bol18}, uses a definition of the hydrostatic bias given by $B = 1/(1-b)$. This combination of cosmological parameters is the only one that can be constrained in the MCMC analysis developed in Sect. \ref{sec:MCMC_planck}. Since the four parameters in the expression of $F$ are completely degenerated, it is not possible to constrain them independently by fitting the tSZ power spectrum alone. The value of $\sigma_8$ can only be estimated by considering auxiliary constraints on the values of $\Omega_m$, $h$ and $b$. This is why many cosmological analyses based on tSZ catalogues or the power spectrum of the tSZ effect also use BAO constraints which bring additional information on the values of $\Omega_m$ and $h$, \emph{e.g.} \citep{pla16c,sal18}. Priors on the value of $b$ are also considered. They are generally based on results from numerical simulations or observations combining the hydrostatic mass of different clusters and mass estimates obtained from weak gravitational lensing \citep[\emph{e.g.}][]{ser17}.

\section{The {\it\bf{{Planck}}} tSZ power spectrum}\label{sec:planck_tsz_power}

The \planck\ satellite enabled obtaining a sky map of the tSZ effect \citep{pla16b}. After masking the regions heavily contaminated by foreground emissions (\emph{e.g.} galactic diffuse emission, infrared and radio sources), this map has been used in order to estimate the power spectrum of the tSZ effect $C_{\ell}^{\mathrm{map}}$. The latter has been computed using the \emph{Xspect} method, developed by \cite{tri05}. This method is based on the crossed power spectra between the sky maps of the tSZ effect obtained by different detectors of the \planck\ satellite. It enables taking into account the masked regions as well as the filtering induced by the \planck\ beam, the analysis of the raw data, and their projection on a pixelated grid. The power spectrum that has been obtained is represented by the grey points in Fig. \ref{fig:planck_power_spec}. The multipole bins have been defined in order to minimize the correlation between adjacent bins at low multipoles and to increase the signal-to-noise at high multipole values \citep{pla16b}. The error bars on the power spectrum $\Delta C_{\ell}^{\mathrm{map}}$ associated with each bin are estimated analytically using the cross-power spectra. They are important at low multipoles due to the sampling variance and start increasing from $\ell \simeq 2000$ due to the spatially correlated residual noise in each map.

The component separation method that has been used in order to obtain the \planck\ $y$-map do not completely exclude all the contaminants to the tSZ signal. Residuals caused by the cosmic infrared background (CIB), unresolved infrared sources and radio sources are still present in the final map used in order to compute the power spectrum of the tSZ effect. The spatially correlated residual noise is also responsible for a significant increase in the power spectrum amplitude measured at high multipoles. The total power spectrum therefore contains several components:
\begin{equation}
C_{\ell}^{\mathrm{map}} = C_{\ell}^{tSZ} + A_{\mathrm{CIB}}\hat{C}_{\ell}^{\mathrm{CIB}} + A_{\mathrm{IR}}\hat{C}_{\ell}^{\mathrm{IR}} + A_{\mathrm{RS}}\hat{C}_{\ell}^{\mathrm{RS}} + A_{\mathrm{CN}}\hat{C}_{\ell}^{\mathrm{CN}}
\label{eq:Cl_model_tot_cosmo}
\end{equation}
where $\hat{C}_{\ell}^{i}$ is the power spectrum associated with the $i^{th}$ component and $A_i$ its amplitude. We consider the tabulated models for the power spectrum of the CIB, the infrared sources (IR), radio sources (RS), and correlated noise (CN) estimated by the \planck\ collaboration \citep{pla16b}. Since the power spectrum of the correlated noise is largely dominant at high multipoles, we use the last measurement point at $\ell = 2742$ in order to estimate the amplitude of its power spectrum. We find that it is given by $A_{\mathrm{CN}} = C_{2742}^{\mathrm{map}}/\hat{C}_{2742}^{\mathrm{CN}} = 0.903$. The amplitudes $A_{\mathrm{CIB}}$, $A_{\mathrm{IR}}$ and $A_{\mathrm{RS}}$ used in Fig. \ref{fig:planck_power_spec} correspond to those estimated by the MCMC analysis developed in Sect. \ref{sec:MCMC_planck}. The contributions of the contaminants in the power spectrum measured by \planck\ are represented by colored lines and their sum corresponds to the black line in Fig. \ref{fig:planck_power_spec}. Contaminants are thus dominant for all multipoles $\ell \gtrsim 1000$. For this reason, we choose to fit the power spectrum measured by \planck\ up to the multipole bin $\ell = 959.5$ in order to optimize the computation time of the analysis while keeping the most constraining points for the power spectrum of the tSZ effect. 

\section{The mean normalized pressure profile}\label{sec:evol_mean_P}

As shown in Eq. (\ref{eq:ClSZ_cosmo_analysis}) and (\ref{eq:tabulated_y}), the mean normalized pressure profile is an essential component of the tSZ power spectrum. In this section, we discuss the properties of the cluster samples that have been used so far in order to estimate the mean normalized pressure profile considered in cosmological analyses. We then explain why the true mean normalized pressure profile of the cluster population could be different from the ones inferred from these samples. The profiles that will be considered in the analysis developed in Sect. \ref{sec:MCMC_planck} are finally defined.

\begin{figure*}
\centering
\includegraphics[height=6.5cm]{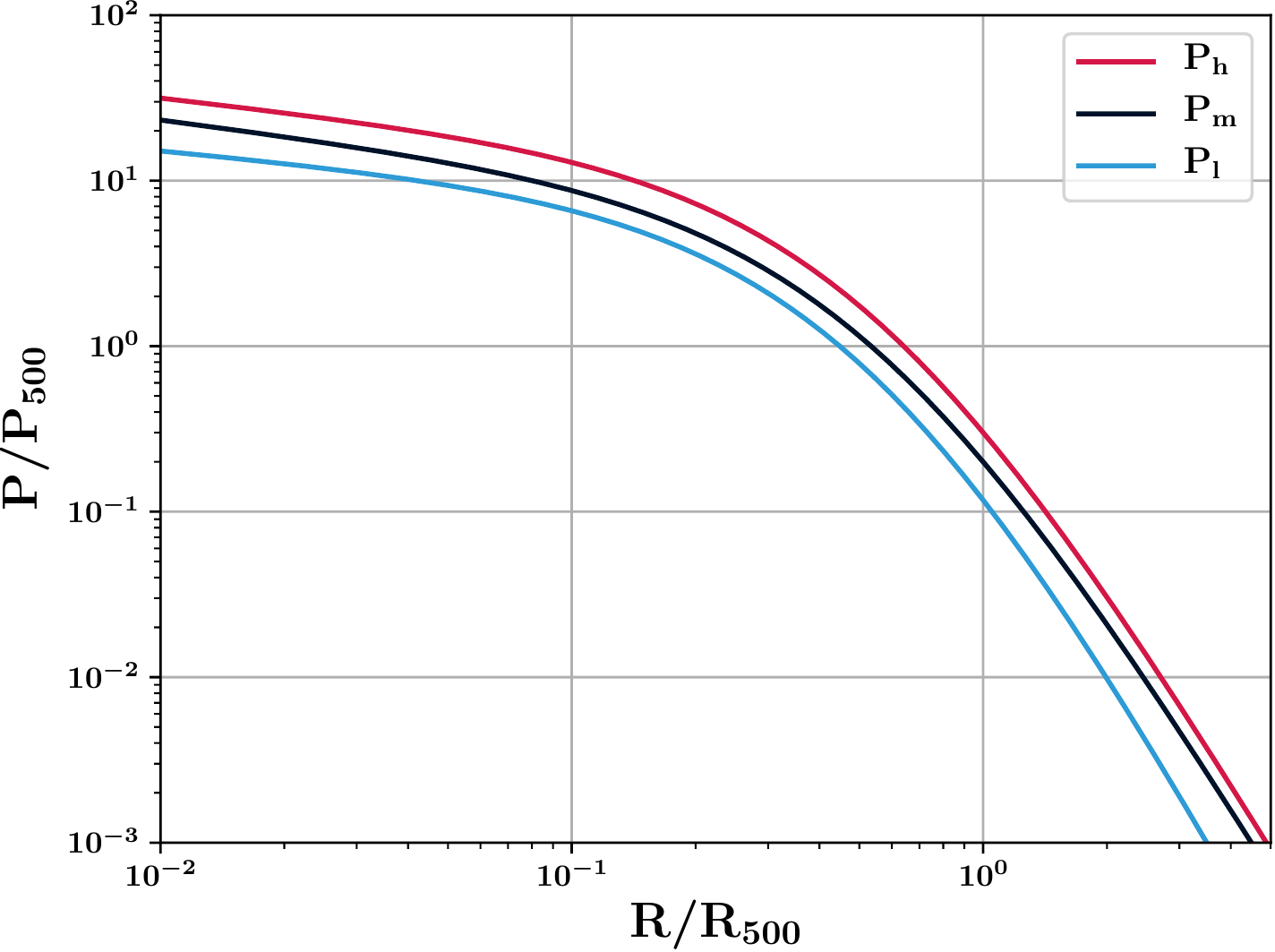}
\hspace{0.2cm}
\includegraphics[height=6.5cm]{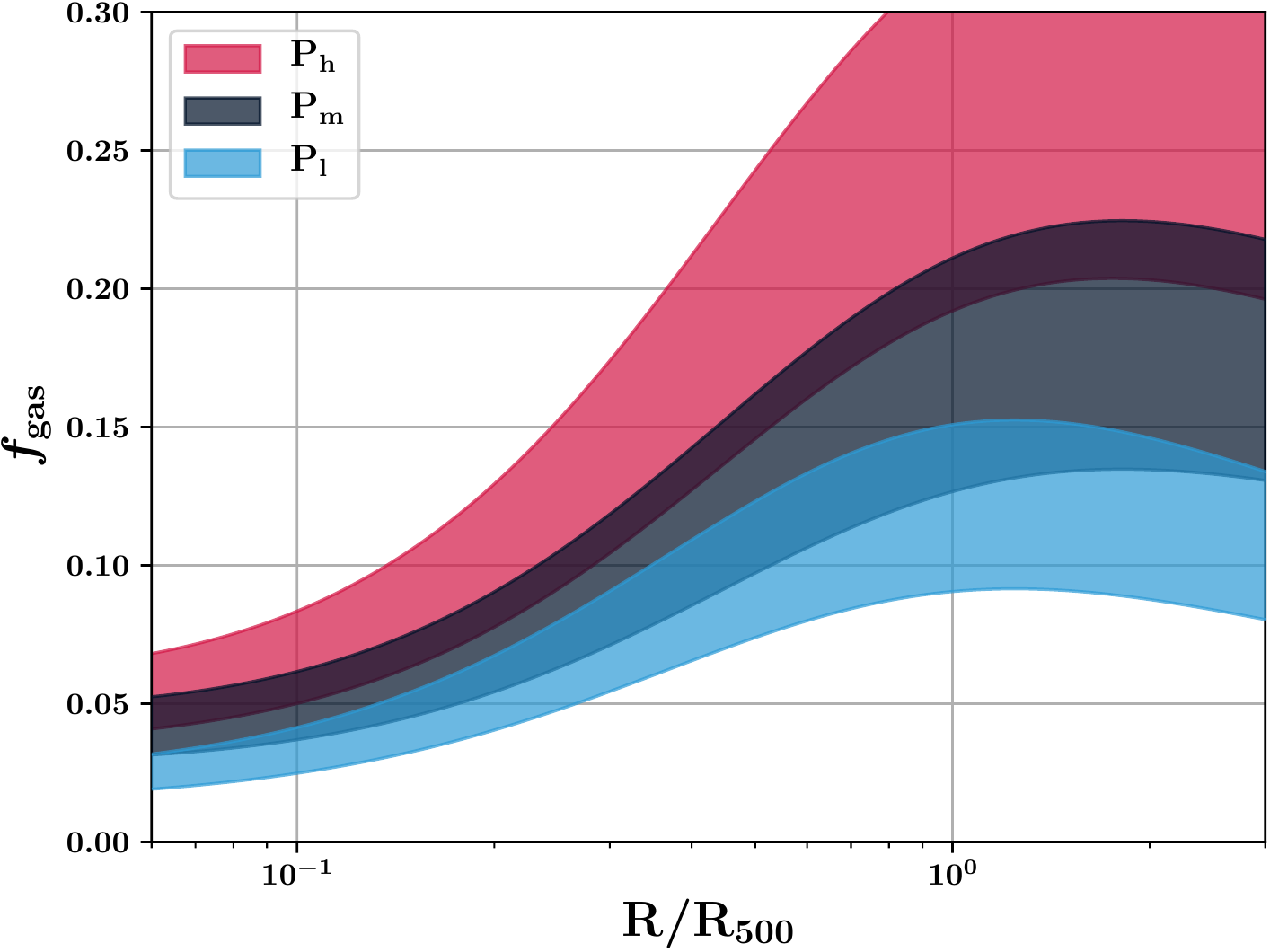}
\caption{\textbf{Left:} Mean normalized pressure profiles considered in the MCMC analysis described in Sect. \ref{sec:MCMC_planck}. The profile obtained by the \planck\ collaboration is shown in black \citep{pla13}. The blue and red profiles are chosen in order to be extreme cases of the profile distribution measured by \planck. \textbf{Right:} Gas mass fraction profiles associated with the mean normalized pressure profiles shown in the left panel (same color code). The width of each profile is due to the uncertainty on the value of the hydrostatic bias parameter. }
\label{fig:P_profiles_gas_frac}
\end{figure*}

\subsection{Current estimates of the mean pressure profile}\label{subsec:current_estimates}

A significant part of the tension observed between the constraints of $\sigma_8$ and $\Omega_m$ estimated from the analysis of the CMB angular power spectrum and the statistical properties of the galaxy cluster population could come from a biased estimate of the mean normalized pressure profile considered in the $I_{\mathds{P}}$ integral (\ref{eq:ClSZ_cosmo_analysis}). We show in Fig. \ref{fig:context_cosmo} the expected cluster abundance for our fiducial cosmology in the mass-redshift plane as well as the REXCESS cluster sample (green symbols), the sub-sample of 62 clusters considered by the \planck\ collaboration \citep{pla13} (purple dots) and the one of the NIKA2 SZ large program \citep{ada18,per18} (orange stars). These cluster samples have been observed or are currently being observed in order to estimate the mean pressure profile used in cosmological analyses. The blue gradient gives the expected number of clusters per unit of mass and redshift defined by:
\begin{equation}
\frac{d^2N}{dz \, dM_{500}} = \int \frac{d^2V}{dz \, d\Omega} \times \frac{dn}{dM_{500}} \, d\Omega
\label{eq:expec_num}
\end{equation}\begin{table*}
\begin{center}
{\footnotesize
\begin{tabular}{cccccc}
\hline
\hline
 & $P_0 \times h_{70}^{-3/2}$ & $c_{\mathrm{500}}$ & $\alpha$ & $\beta$ & $\gamma$ \\
\hline
$P_h$ &  10.20 & 1.80  & 1.33  & 4.20 & 0.27  \\
$P_m$ &  6.41 & 1.81  & 1.33  & 4.13 & 0.31 \\
$P_l$ & 6.00 & 1.80  & 1.30  & 4.60 & 0.22  \\
\hline
\hline
\end{tabular}}
\end{center}
\caption{Parameters of the gNFW pressure profile models shown in Fig. \ref{fig:P_profiles_gas_frac} and considered in the cosmological analysis described in Sect. \ref{sec:MCMC_planck}.}
\label{tab:param_gnfw_cosmo}
\end{table*}where $d^2V/dzd\Omega$ is the comoving volume element for a solid angle $d\Omega$ and $dn/dM_{500}$ is the mass function giving the number density of halos. Although the amplitude of the tSZ effect induced by individual low-mass halos is lower than the one caused by massive clusters (see Eq. (\ref{eq:P500})), their high number density explains why their contributions to the overall tSZ power spectrum measured by \planck\ is dominant \citep[see \emph{e.g.}][]{pla16b}. However, the mean normalized pressure profiles that are most often considered in cosmological analyses based on tSZ surveys are the ones estimated by the \planck\ and REXCESS samples \citep{pla13,pra09}. These samples both contain exclusively massive and low redshift ($z < 0.5$) clusters because observations of high-redshift or low-mass clusters are still very challenging.\\
If the whole cluster population is not exactly self-similar, the distribution of pressure profiles observed at low mass and high redshift will be different from the ones measured from the analysis of the \planck\ and REXCESS samples. Selection biases due to wrong estimations of noise properties or observational biases toward a given cluster morphology may also induce such differences. It is therefore essential to estimate the impact of a potential modification of the mean normalized pressure profile on the cosmological constraints established from the analysis of tSZ survey data.

\subsection{Mean pressure profile of the cluster population}\label{subsec:selection_P_prof}

We consider the distribution of pressure profiles measured by the \planck\ collaboration on 62 nearby clusters in order to study the impact of the shape and amplitude of the mean normalized pressure profile on the estimation of $\sigma_8$ and $\Omega_m$ from the study of the tSZ power spectrum \citep{pla16b}. As shown in Fig. \ref{fig:context_cosmo}, this sample\footnote{like any sample selected in a limited range of mass and redshift} is not representative of the total population of galaxy clusters. If the assumption of galaxy cluster self-similarity is not verified at high redshift or for masses lower than $3\times 10^{14}~\mathrm{M_{\odot}}$, the mean pressure profile associated with the cluster population could be significantly different from the one estimated with this sample of nearby clusters. This mean profile called $P_m$ is represented by a black line in Fig. \ref{fig:P_profiles_gas_frac}.\\

Two pressure profiles are defined on either side of the $P_m$ profile using the gNFW parametric model given by Eq. (\ref{eq:cosmo_normalized_P}). These profiles are chosen in order to define two extreme cases of the observed distribution of normalized pressure profiles. We name $P_l$ and $P_h$ the normalized pressure profiles shown in Fig. \ref{fig:P_profiles_gas_frac} in the case where the mean ICM pressure distribution of the cluster population is respectively lower and higher than the mean profile estimated by the analysis of the \planck\ cluster sample. The parameters of the $P_l$ and $P_h$ profiles are obtained combining the requirements of the profiles being inside the intrinsic scatter of the distribution of normalized pressure profiles obtained by \planck\ \citep{pla13} and that the corresponding gas mass fraction is compatible with the observed values at low redshift. The methodology used in order to estimate the gas mass fraction profile for a given pressure profile $P(r)$ is described bellow.\\

Let a galaxy cluster of redshift $z$ and mass $M_{500}$. Knowing the mass and redshift of the cluster, we estimate the characteristic radius $R_{500}$ for our fiducial cosmological model:
\begin{equation}
        R_{500} = \left[ \frac{3 M_{500}}{4\pi \, 500\rho_c(z)}\right]^{1/3}
\label{eq:R500}
\end{equation}
We assume that the cluster mass profile is given by a Navarro-Frenk-White (NFW) model \citep{nav97}:
\begin{equation}
        M_{\mathrm{tot}}(r) = 4\pi \, r_s^3 \, \rho_0 \left[ \mathrm{ln}\left(\frac{r_s + r}{r_s}\right) - \frac{r}{r_s+r}\right]
\label{eq:Mtot}
\end{equation}
Following the results shown in Fig. 15 of \cite{miy18}, we assume that the concentration parameter of the mass profile is given by $c_{500}^{NFW} = 2.5$. The characteristic radius $r_s$ in the NFW model is therefore given by:
\begin{equation}
        r_s= \frac{R_{500}}{c_{500}^{NFW}}
\label{eq:rs}
\end{equation}
Knowing the values of $r_s$, $R_{500}$, and $M_{500}$, we deduce the one of the amplitude parameter $\rho_0$ in Eq. (\ref{eq:Mtot}). In this study, we assume that the hydrostatic bias is compatible with the current constraints \citep[see \emph{e.g.} Fig. 10 in][]{sal18}. The value of $b$ is thus defined by a uniform distribution bounded between 0 and $0.4$. Furthermore, we assume that the radial profile of the hydrostatic bias parameter is constant. We estimate the hydrostatic mass profile of galaxy clusters based on the known NFW model and the considered value of $b$:
\begin{equation}
        M_{\mathrm{HSE}}(r)= (1-b) M_{\mathrm{tot}}(r)
\label{eq:Mhse}
\end{equation}
Knowing both the ICM pressure profile given by Eq. (\ref{eq:cosmo_P_prof}) and the hydrostatic mass profile given by Eq. (\ref{eq:Mhse}), we deduce the ICM density profile:
\begin{equation}
        n_e(r)= \frac{-r^2}{G \mu m_p M_{\mathrm{HSE}}(r)} \frac{dP(r)}{dr}
\label{eq:density}
\end{equation}
where $G$ is the Newton's gravitational constant, $m_p$ the proton mass, and $\mu=0.6$ is the gas mean molecular weight. The spherical integral of the ICM density profile allows us to obtain the cluster gas mass profile:
\begin{equation}
        M_{\mathrm{gas}}(r)= 4\pi \, \mu_e m_p \, \int_0^r n_e(r') \, r'^2 \, dr'
\label{eq:Mgas}
\end{equation}
where $\mu_e = 1.15$ is the electron mean molecular weight. The gas mass fraction profile is finally computed from the ratio of the gas mass profile (\ref{eq:Mgas}) and the total mass profile of the cluster (\ref{eq:Mtot}):
\begin{equation}
        f_{\mathrm{gas}}(r)= \frac{M_{\mathrm{gas}}(r)}{M_{\mathrm{tot}}(r)}
\label{eq:fgas}
\end{equation}
The amplitude of this profile as a function of $r/R_{500}$ does not depend on the values of $z$ and $M_{500}$ but it strongly depends on the considered hydrostatic bias. We therefore compute the gas mass fraction profiles under the assumption that $b=0$ and $b=0.4$ for each mean normalized  pressure profile that we consider. One mean normalized pressure profile is therefore associated with an interval of gas mass fraction profiles with a lower bound corresponding to $b=0$ and an upper bound obtained with $b=0.4$. The $P_l$ and $P_h$ profiles are defined so that they are compatible with the distribution of pressure profiles measured at low redshift by \planck\ \citep{pla13} and under the additional constraint that their corresponding gas mass fraction profiles are compatible with the measured distributions given in \cite{pla13,eck13,eck19}. They are represented in the left panel of Fig. \ref{fig:P_profiles_gas_frac} in blue and red respectively. Their corresponding gas mass fraction profiles are contained within the colored interval represented in the right panel of Fig. \ref{fig:P_profiles_gas_frac}. We emphasize that each interval is only associated with one mean normalized profile shown in the left panel. The width of these interval is only due to the uncertainty on the value of the hydrostatic bias parameter. The parameters of the three mean normalized pressure profiles shown in Fig. \ref{fig:P_profiles_gas_frac} and used in the analysis developed in Sect. \ref{sec:MCMC_planck} are summarized in table \ref{tab:param_gnfw_cosmo}.

\subsection{Scaling parameter}\label{subsec:scaling_rel}

\begin{figure*}
\centering
\includegraphics[height=7.7cm]{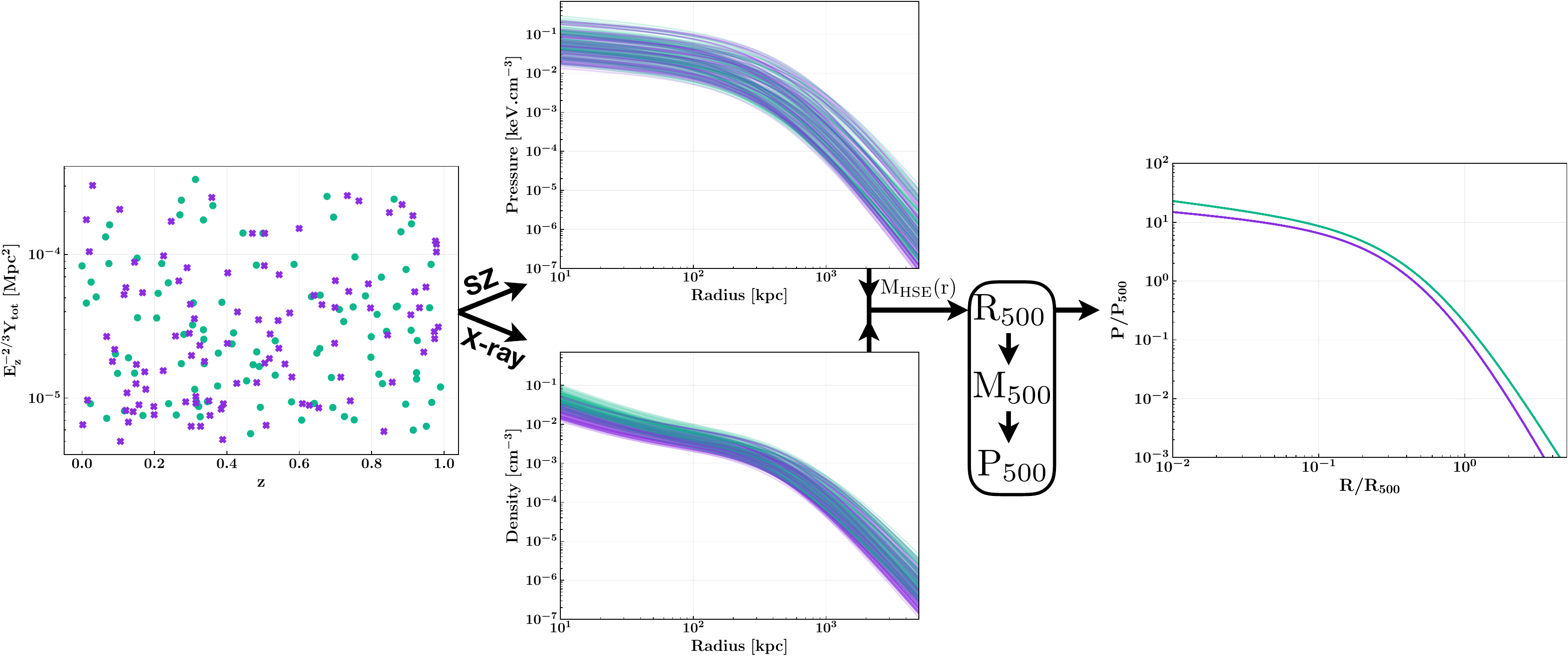}
\caption{\textbf{Left:} Two simulated samples of 100 clusters with identical distributions in integrated Compton parameter and redshift. \textbf{Center:} Pressure profiles (up) and density profiles (down) associated with the two simulated samples. The sample shown in green (magenta) has a mean gas mass fraction profile corresponding to the one associated with the $P_m$ ($P_l$) profile (see Fig. \ref{fig:P_profiles_gas_frac}). The scaling parameters $P_{500}$ for all the simulated profiles are estimated using the hydrostatic equilibrium assumption and Eq. \ref{eq:P500}. \textbf{Right:} Mean pressure profiles associated with each sample after normalization of all the simulated pressure profiles by their corresponding scaling parameter.}
\label{fig:simu_scaling}
\end{figure*}
\begin{table*}
\begin{center}
{\footnotesize
\begin{tabular}{ccc}
\hline
\hline
Parameters & Min & Max \\
\hline
$F$ & 0.2 & $\infty$\\
$\Omega_m$ & 0.1 & 1.0\\
$b$ & 0 & 0.4\\
$A_{\mathrm{CIB}}$ & 0 & 10\\
$A_{\mathrm{IR}}$ & 0 & 10\\
$A_{\mathrm{RS}}$ & 0 & 10\\
\hline
\hline
\end{tabular}}
\end{center}
\caption{Interval boundaries defining the uniform priors associated with the free parameters listed in the left column and used in the MCMC analysis described in Sect. \ref{sec:MCMC_planck}.}
\label{tab:priors_cosmo}
\end{table*}

\begin{figure*}
\centering
\includegraphics[height=14cm]{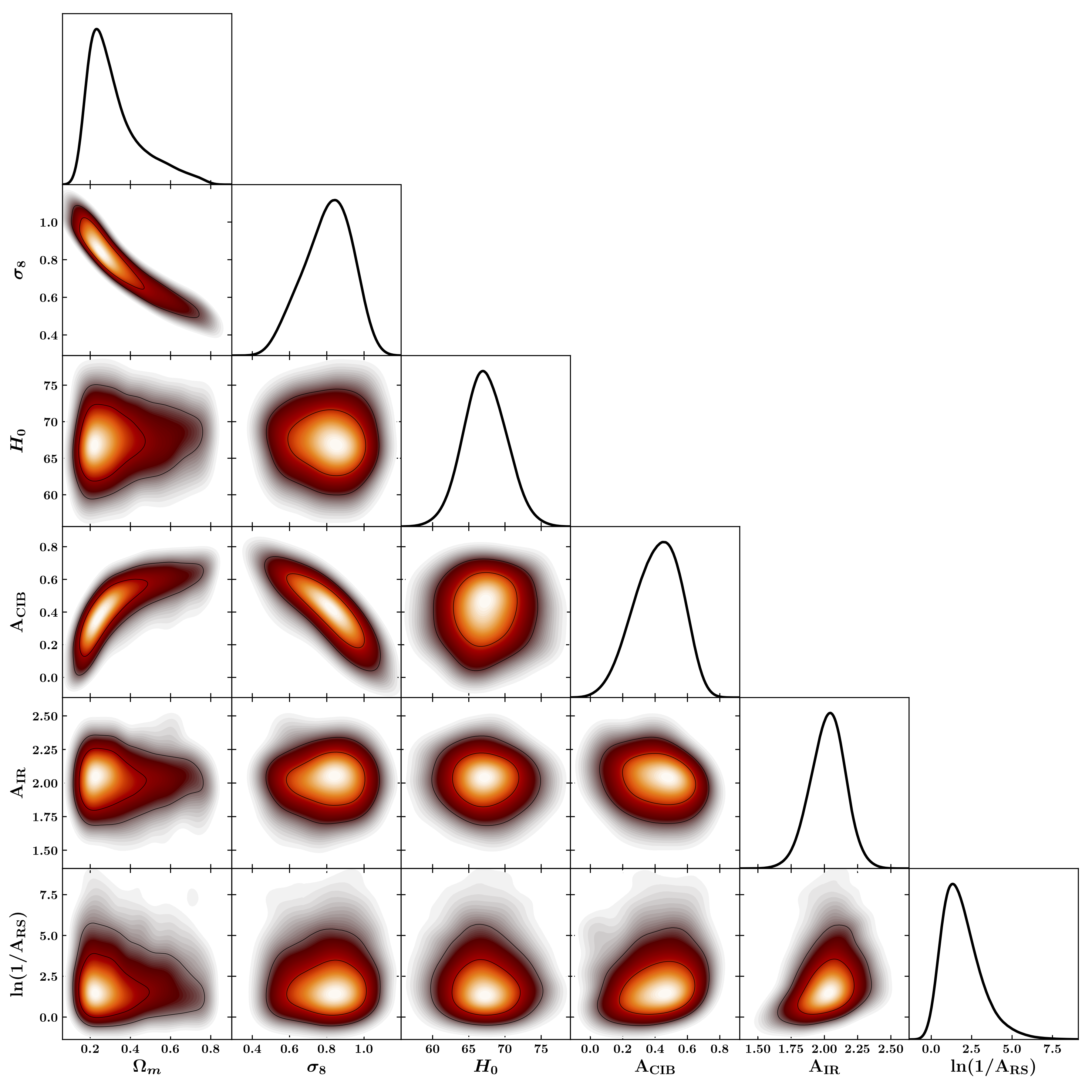}
\caption{Marginalized distributions (diagonal) and 2D correlations (off-diagonal) plots of the $\Omega_m$, $\sigma_8$, and $H_0$ cosmological parameters along with the amplitudes of the astrophysical contaminants to the tSZ signal obtained at the end of the MCMC analysis based on the $P_m$ pressure profile (see Sect. \ref{sec:MCMC_planck}).}
\label{fig:contours_MCMC}
\end{figure*}

The profiles $P_l$, $P_m$ and $P_h$ are used in independent analyses in Sect. \ref{sec:MCMC_planck} to constrain the $F$ parameter with the tSZ power spectrum measured by \planck. Another key parameter used in the definition of the tSZ power spectrum in Eq. (\ref{eq:ClSZ_cosmo_analysis}) is the scaling parameter $P_{500}$. As shown in Eq. (\ref{eq:P500}), it depends on the mass and redshift of galaxy clusters. The product of the scaling parameter associated with a given cluster with the mean normalized pressure profile of the cluster population enables retrieving its true pressure distribution and, in turn, its expected tSZ signal. For this reason, it is essential to use the same definition of the scaling parameter for the normalization of the measured pressure profiles in a given sample of galaxy clusters and for the cosmological analysis based on the corresponding mean normalized pressure profile.\\

We have realized a simulation to show how a modification of the mean gas mass fraction profile of the cluster population has an impact on the values of the scaling parameters $P_{500}$ estimated for a sample of galaxy clusters. We show in the left panel of Fig. \ref{fig:simu_scaling} two different samples of 100 clusters simulated at redshifts $0 < z < 1$ based on the ICM model described in Sect. \ref{subsec:model_tsz_power} and \ref{subsec:selection_P_prof}. The distributions of the total tSZ signal and redshift have been computed to be uniform along both axes. We simulate the pressure and density profiles of the sample shown in green (central panel) based on the mean normalized pressure profile $P_m$ and the gas mass fraction profile shown in grey in Fig. \ref{fig:P_profiles_gas_frac} for a hydrostatic bias $b=0.2$. These two profile distributions are used to compute the associated hydrostatic mass profiles using Eq. (\ref{eq:density}). These mass profiles are then used to estimate the values of $R_{500}$ which allow us to obtain the mass of each cluster based on Eq. (\ref{eq:R500}). Knowing these masses, we apply the definition of the scaling parameter $P_{500}$ given in Eq. (\ref{eq:P500}) in order to obtain the mean normalized pressure profile shown in green in the right panel of Fig. \ref{fig:simu_scaling}. The same methodology is applied for the sample in purple. However, for this sample, we assume the mean normalized pressure profile $P_l$ and a lower gas mass fraction given by the blue profile in Fig. \ref{fig:P_profiles_gas_frac} to simulate the pressure and density profiles. In this situation, the distribution of pressure profiles associated with the selected sample is very similar to the one obtained with the green cluster sample. However, the decrease in gas mass fraction results in a slight decrease of the amplitude of the density profiles. The hydrostatic mass profiles computed from Eq. (\ref{eq:density}) are consequently higher than the ones obtained with the green profiles. This leads to larger values of the scaling parameter $P_{500}$ computed for each cluster from the definition given in Eq. (\ref{eq:P500}). The resulting mean normalized pressure profile shown in purple in the right panel of Fig. \ref{fig:simu_scaling} is therefore lower than the one associated with a higher gas mass fraction.\\

The important point to notice here is that the same definition of the scaling parameter given in Eq. (\ref{eq:P500}) has been applied although the gas mass fraction is different in both samples. As it is reminded by \cite{arn10} in Appendix A, the exact choice of the numerical coefficients in Eq. (\ref{eq:P500}) does not need to be consistent with the expected values from the self-similar model. It is however essential to use the same definition of the scaling parameter for the calibration of the mean normalized pressure profile and for the cosmological analyses that rely on it.\\

We note that the simulated cases shown in Fig. \ref{fig:simu_scaling} are just examples of a way to measure different mean normalized pressure profiles in a given range of redshift and integrated Compton parameter. For example, the intrinsic scatter of the pressure profile distribution associated with the total cluster population may also be different from the one observed at low redshift. A mean normalized pressure profile lower than $P_m$ could therefore be obtained if we measure an asymmetric distribution of normalized pressure profiles with a high density of profiles of low amplitude. In this paper, we will assume that the three profiles $P_l$, $P_m$ and $P_h$ have been obtained using a methodology that is similar to the one displayed in Fig. \ref{fig:simu_scaling}. We will therefore apply the same definition of the scaling parameter given in Eq. (\ref{eq:P500}) for the three profiles. This allows us to study the effect of the mean pressure profile measured from the study of representative cluster samples on the cosmological parameters obtained from the analysis of tSZ survey data. This is important in the context of forthcoming studies such as the NIKA2 tSZ large program \citep[see][for more details on this program]{com16,per18}.

\section{Analysis of the {\it\bf{{Planck}}} tSZ power spectrum}\label{sec:MCMC_planck}

The three normalized pressure profiles shown in Fig. \ref{fig:P_profiles_gas_frac} are used in order to tabulate the values of the associated integrals $I_{\mathds{P}}$ (see Eq. (\ref{eq:tabulated_y})). A MCMC procedure is used to sample the space of cosmological parameters and thus constrain the combination of parameters $F$ (see Sect. \ref{subsec:cosmo_tsz_power}) by fitting the power spectrum $C_{\ell}^{\mathrm{map}}$ measured by \planck\ \citep{pla16b}. The parameters that are left free in the analysis are $F$, $\Omega_m$, $b$, $h$, $A_{\mathrm{CIB}}$, $A_{\mathrm{IR}}$ and $A_{\mathrm{RS}}$. The amplitude $A_{\mathrm{CN}}$ keeps a fixed value of 0.903 (see Sect. \ref{sec:planck_tsz_power}). At each step of the MCMC, the equation (\ref{eq:ClSZ_cosmo_analysis}) is used to compute the tSZ power spectrum for the current cosmological model and the considered pressure profile. The amplitudes of the power spectra associated with the contaminants are used in order to calculate the total power spectrum model $\hat{C}_{\ell}^{\mathrm{tot}}$ by applying Eq. (\ref{eq:Cl_model_tot_cosmo}). The total power spectrum is then compared to the power spectrum measured by \planck\ for the 18 bins constrained at multipoles below 1000 using the Gaussian likelihood function $\mathscr{L}$ defined by:
\begin{equation}
-2 \mathrm{ln} \, \mathscr{L} = \sum_{\ell} \left[\frac{C_{\ell}^{\mathrm{map}} - \hat{C}_{\ell}^{\mathrm{tot}}}{\Delta C_{\ell}^{\mathrm{map}}}\right]^2
\end{equation}
\begin{figure*}
\centering
\includegraphics[height=7cm]{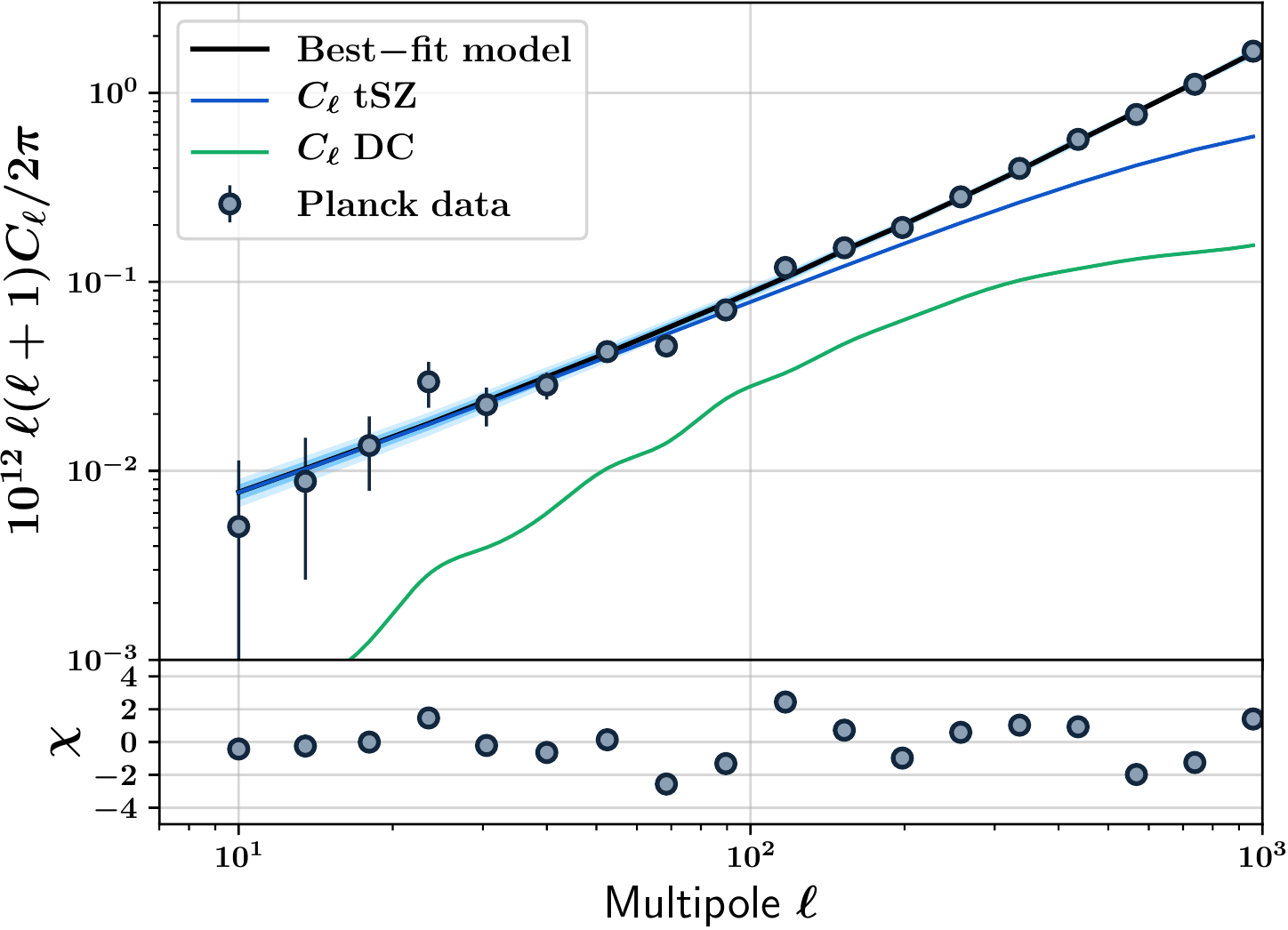}
\caption{Best fit model of the \planck\ tSZ power spectrum (black line) obtained with the analysis based on the $P_m$ profile and associated 1 and $2\sigma$ uncertainties (dark and light blue regions). The considered data-set is represented with the grey points. The component of the total power spectrum that gives the tSZ power spectrum (see Eq. \ref{eq:ClSZ_cosmo_analysis}) is shown with the blue line. The tSZ power spectrum associated with the detected clusters in the PSZ2 catalog is also shown in green \citep{pla16}. The lower panel gives the statistical significance in $\sigma$ units of the residuals obtained by subtracting the best-fit model to the data.}
\label{fig:best_fit_model}
\end{figure*}
\begin{table*}
\begin{center}
{\footnotesize
\begin{tabular}{cc}
\hline
\hline
 & $F$ \\
 \hline
P high & $0.437\pm 0.006$ \\
P mean & $0.484 \pm 0.005$\\
P low & $0.530 \pm 0.008$\\
Bolliet \emph{et al.} & $0.473 \pm 0.005$\\
\planck\ XXII & $0.477 \pm 0.004$\\
CMB+BAO & $0.520 \pm 0.013$\\
\hline
\hline
\end{tabular}}
\end{center}
\caption{Constraints of the $F$ parameter, cf. eq (\ref{eq:defintion_F}), obtained from the three considered pressure profiles. We also show the results given by the tSZ power spectrum analyses realized by \citep{bol18} and \citep{pla16b}. The last estimate is computed from the cosmological constraints obtained from the analysis combining the \planck\ CMB power spectrum and BAO data \citep{pla18}. All the error bars represent the 68\% confidence level on the best-fit values.}
\label{tab:results_F}
\end{table*}

The error bars $\Delta C_{\ell}^{\mathrm{map}}$ associated with the points measured by \planck\ \citep{pla16b} do not take into account the impact of the trispectrum\footnote{See \cite{bol18} for more details on the trispectrum.} in the analysis of the power spectrum of the tSZ effect. Although this contribution is not negligible for $\ell \lesssim 200$, we did not consider it for computing-time reasons. Uniform priors are considered for each free parameter except $h$. A Gaussian prior is considered for the parameter $h = 0.67\pm 0.03$ according to the estimate obtained by the \planck\ CMB analysis \citep{pla18}. The boundaries of the parameter priors are defined in table \ref{tab:priors_cosmo}.

The $y$-map used by the \planck\ collaboration in order to estimate the tSZ power spectrum also contains the tSZ signal caused by all clusters with a signal-to-noise ratio high enough to be detected. The knowledge of the position and integrated tSZ signal of the clusters listed in the PSZ2 catalog \citep{pla16} thus makes it possible to estimate the power spectrum of the tSZ effect associated with these detected clusters $C_{\ell}^{\mathrm{DC}}$ \citep{pla16b}. The sum of the power spectra associated with the contaminants and the detected clusters cannot exceed the total power spectrum measured by \planck. Following the methodology presented in \citep{bol18}, an additional constraint on the amplitudes of the power spectra of the contaminants is thus used at each step of the MCMC:
\begin{equation}
A_{\mathrm{CIB}}\hat{C}_{\ell}^{\mathrm{CIB}} + A_{\mathrm{IR}}\hat{C}_{\ell}^{\mathrm{IR}} + A_{\mathrm{RS}}\hat{C}_{\ell}^{\mathrm{RS}} + A_{\mathrm{CN}}\hat{C}_{\ell}^{\mathrm{CN}} + C_{\ell}^{\mathrm{DC}} < C_{\ell}^{\mathrm{map}}
\end{equation}
\begin{figure*}
\centering
\includegraphics[height=7.5cm]{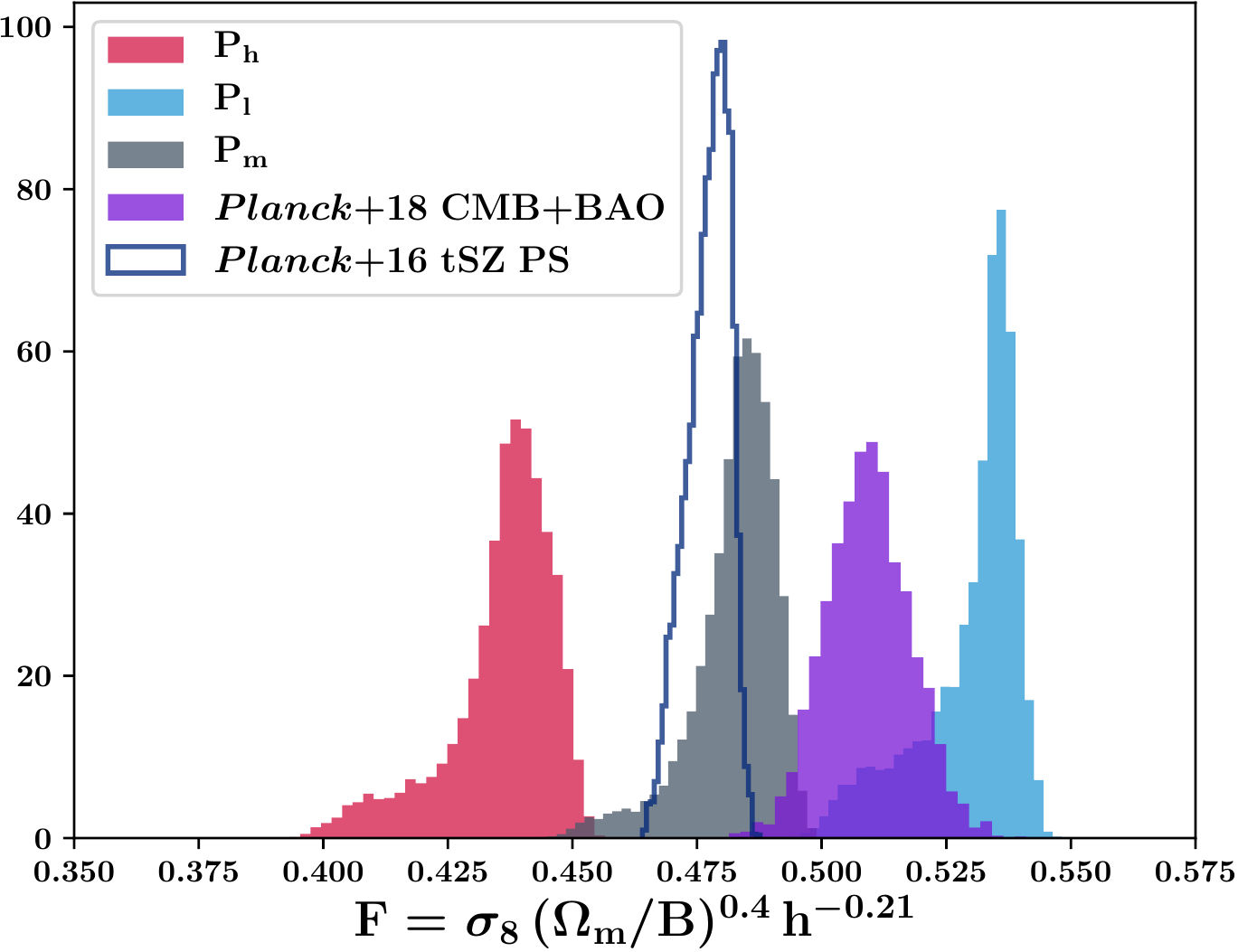}
\caption{Distributions of the combination of cosmological parameters $F$ obtained with the three MCMC analyses based on the $P_h$ (red), $P_m$ (grey), and $P_l$ (blue) pressure profiles. The distribution estimated by the \planck\ collaboration from an analysis based on the $P_m$ profile, for a hydrostatic bias fixed to $b=0.2$, and a different prior on $H_0$ is shown with the dark blue line. The distribution of $F$ obtained from the joint analysis of the CMB primary anisotropies and BAO data is also shown in purple for a fixed hydrostatic bias $b=0.2$.}
\label{fig:results_no_prior}
\end{figure*}
A total of 240 Markov chains is used to explore the parameter space efficiently. The convergence test of \cite{gel92} is used in order to stop the MCMC sampling and we also compute the chain autocorrelation functions to select the final independent samples in the posterior distribution. A \emph{burn-in} cut-off that excludes the first half of the samples is applied to each chain in order to ensure the independence of the stored samples from the initial position in the parameter space. The remaining samples are used to estimate the marginalized probability densities associated with each parameter. This analysis is carried out independently for the three mean normalized pressure profiles defined in Sect. \ref{subsec:selection_P_prof}.

We show in Fig. \ref{fig:contours_MCMC} the one- and two-dimensional marginalized distributions of the parameters of interest fitted using the normalized pressure profile $P_m$. We observe that the parameters $\sigma_8$ and $\Omega_m$ are clearly anti-correlated. They are also both correlated with the $H_0$ parameter. These correlations are expected (see Sect. \ref{subsec:cosmo_tsz_power}) and justify \emph{a posteriori} the definition of the $F$ parameter characterizing the amplitude of the measured power spectrum. The $F$ parameter being the only one that can be constrained by the analysis of the tSZ power spectrum, the $\sigma_8$ and $\Omega_m$ parameters are completely degenerated. The priors associated with the parameters of interest being uniform, the contours expected in the $\sigma_8$-$\Omega_m$ plane correspond to a diagonal strip of negative slope, with a central position constrained by the data but with a length entirely defined by the boundaries of the priors. This explains why the $\Omega_m$ parameter is scattered between 0.1 and 1 in Fig. \ref{fig:contours_MCMC}.\\

The one- and two-dimensional marginalized distributions of the amplitudes of the contaminants are also represented in Fig. \ref{fig:contours_MCMC}. The estimated amplitudes are compatible with those obtained in the two other analyses based on the $P_l$ and $P_h$ profiles. The estimates  
\begin{equation}
A_{\mathrm{CIB}} = 0.38 \pm 0.14, A_{\mathrm{IR}} = 2.02 \pm 0.13~ \mathrm{and} ~A_{\mathrm{RS}} = 0.26^{+0.37}_{-0.16}
\end{equation}
are also compatible with the constraints established by \cite{bol18} using the same data set. We show in Fig. \ref{fig:A1_compare} a full comparison between the constraints on cosmological and nuisance parameters when the three profiles $P_l$, $P_m$, and $P_h$ are considered. The only effect of a modification of the mean normalized pressure profile on the posterior distribution is a variation of the best-fit value of the $\sigma_8$ parameter.\\

As tSZ data alone cannot be used to constrain the hydrostatic bias parameter, the posterior distributions associated with this nuisance parameter are all compatible with the prior uniform distributions between 0 and 0.4.\\

The marginalized distributions of the $\sigma_8$, $\Omega_m$, $b$, and $h$ parameters are used to compute the tSZ power spectrum model at the maximum likelihood. The power spectrum model obtained at the end of the analysis based on the $P_m$ profile is represented by a black line in Fig. \ref{fig:best_fit_model}. The uncertainties at 1 and $2\sigma$ associated with the best-fit model are estimated by Monte Carlo drawing of parameter sets computed from the Markov chains. They are represented by the dark blue and light blue regions. They are only visible at multipoles smaller than 100 where the sampling variance limits the accuracy of the measurements\footnote{We note that they would be more important if we considered the impact of the trispectrum in the measurement errors.}. The lower panel in Fig. \ref{fig:best_fit_model} represents the residual statistical significance after subtracting the best fit to the measured power spectrum. No residual greater than $3\sigma$ is observed, which support the good accuracy of the fit. The same conclusion is drawn from the analyses based on the $P_l$ and $P_h$ profiles. We show the corresponding results in Fig. \ref{fig:A2_bestfits} and do not find any significant difference with the ones obtained with the $P_m$ profile.\\

The blue line in Fig. \ref{fig:best_fit_model} corresponds to the tSZ power spectrum contribution to the total power spectrum. The green line represents the power spectrum of the tSZ effect associated with the detected clusters, listed in the PSZ2 catalog \citep{pla16}. This comparison highlights the fact that most of the tSZ signal contained in the \planck\ $y$-map is caused by low mass clusters that have not been detected but whose abundance is much larger than the one of the detected clusters.

\section{Implication of the mean pressure profile on the cosmological constraints}\label{sec:results}

\begin{figure*}
\centering
\includegraphics[height=7.9cm]{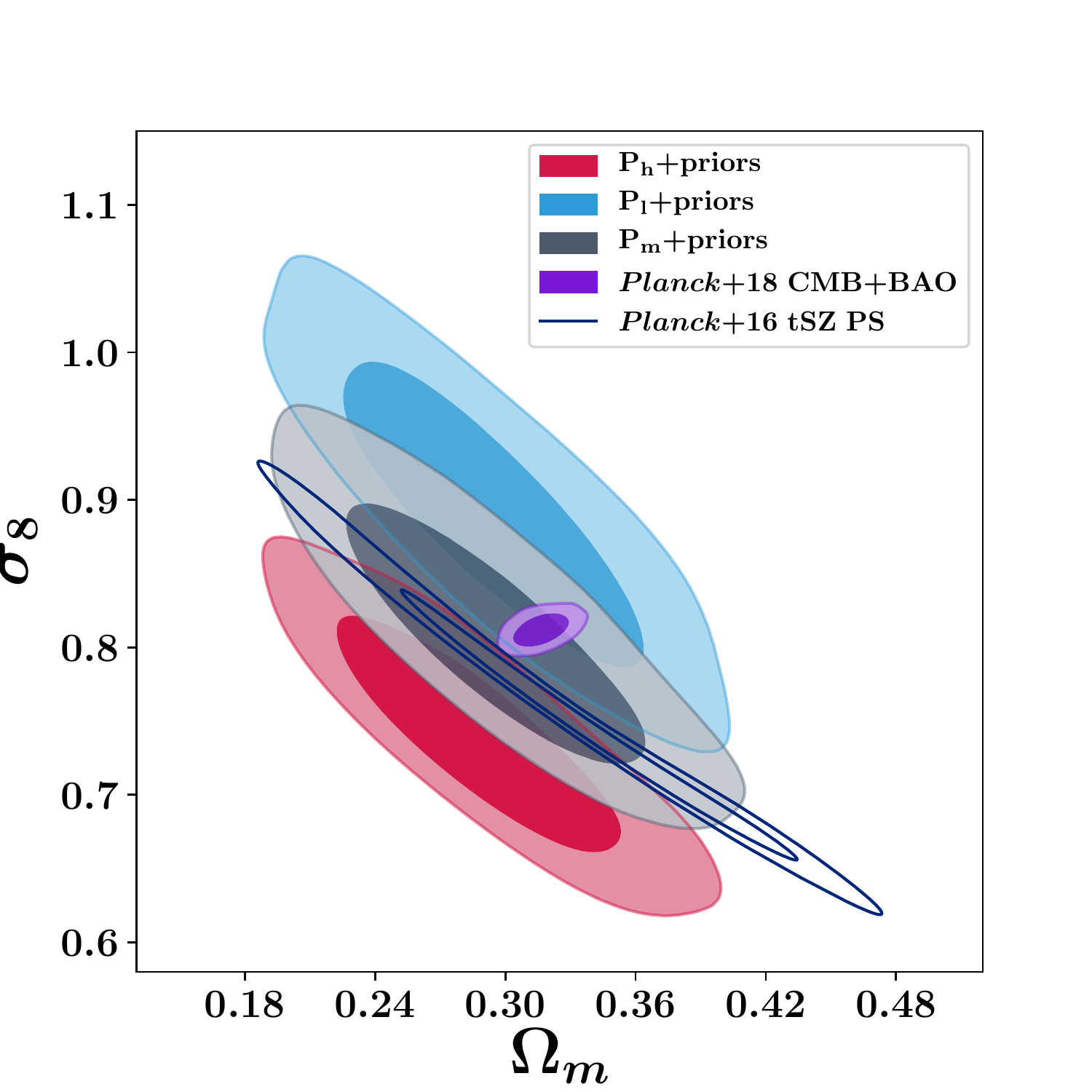}
\hspace{0.4cm}
\includegraphics[height=7cm]{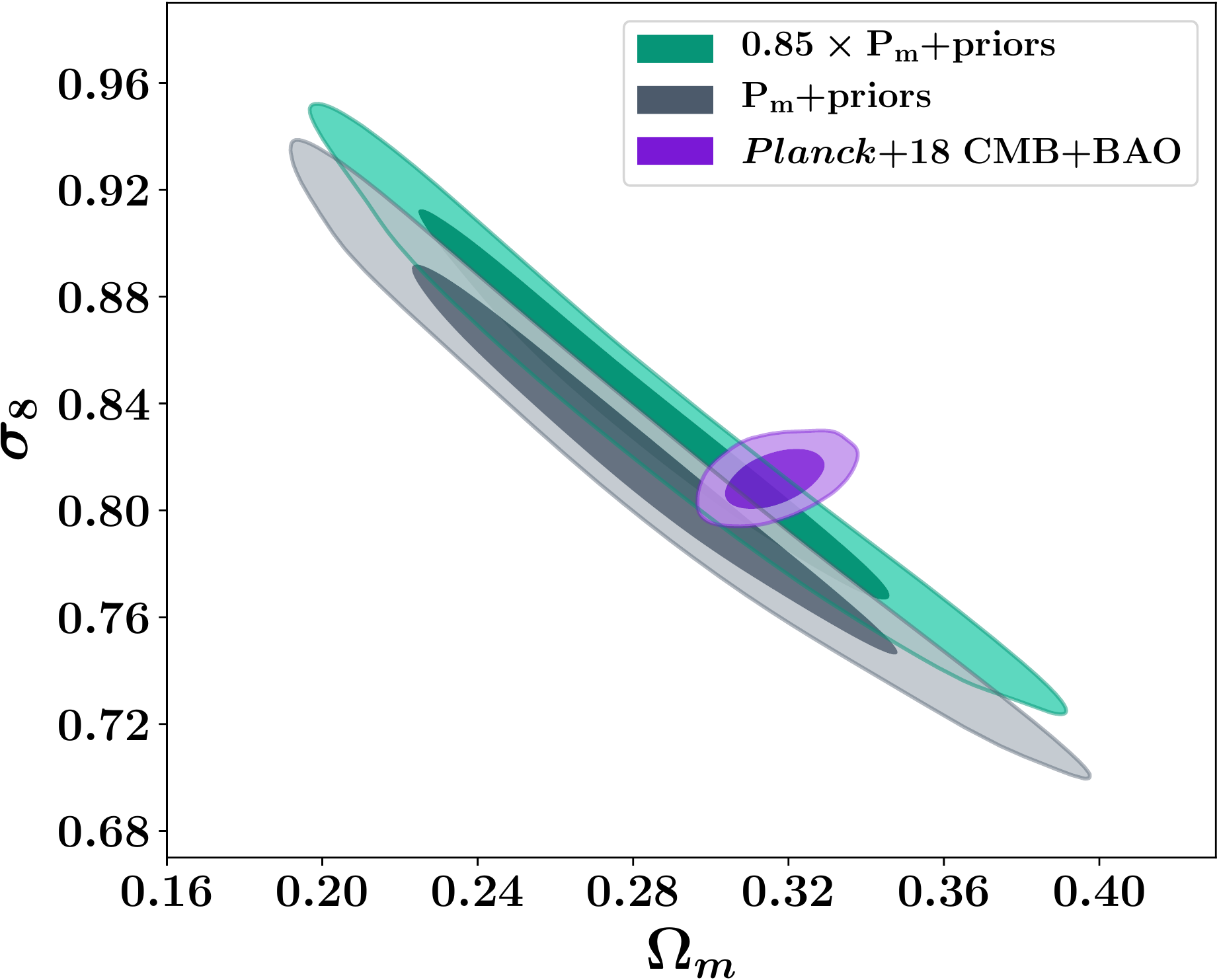}
\caption{\textbf{Left:} Constraints obtained on the $\sigma_8$ and $\Omega_m$ cosmological parameters issued from the three MCMC analyses based on the $P_h$ (red), $P_m$ (grey), and $P_l$ (blue) pressure profiles including informative priors on the values of $\Omega_m$ and $b$ (see Sect. \ref{sec:results}). The constraints obtained from the joint analysis of the CMB primary anisotropies and BAO data and from the \planck\ analysis of the tSZ power spectrum for a fixed hydrostatic bias $b=0.2$ are also shown in purple and dark blue respectively. \textbf{Right:} Constraints obtained on the $\sigma_8$ and $\Omega_m$ cosmological parameters using the $P_m$ profile (grey) and the $P_m$ profile scaled down by 15\% (green) for a possible future hydrostatic bias prior of $b = 0.20 \pm 0.01$. The CMB contours (purple) are the same as the ones shown in the left panel. Note that the range of the y-axis is different.}
\label{fig:results_prior}
\end{figure*}

The distributions of the $F$-parameter estimates are shown in Fig. \ref{fig:results_no_prior} for the profiles $P_h$ (red), $P_m$ (grey) and $P_l$ (blue). We compare the distribution of $F$ obtained for the $P_m$ profile with the one estimated from the analysis of the tSZ power spectrum performed by the \planck\ collaboration \citep{pla16b} based on the same profile (dark blue line). The maximum likelihood positions for the two distributions are slightly different because we have used a different prior for the Hubble parameter $h$. In addition, the width of the $F$ distribution associated with the $P_m$ profile is larger than the one associated with the distribution constrained by the \planck\ analysis. Indeed, the latter was based on a hydrostatic bias parameter fixed to a value of 0.2 unlike the analysis developed in this paper where $b$ varies according to a uniform distribution centered on 0.2 (see table \ref{tab:priors_cosmo}). The constraints from the three analyses are listed in table \ref{tab:results_F}. The fact that the estimate of $F$ obtained through the analysis based on the $P_m$ profile developed in this paper is compatible with the ones obtained by \cite{bol18} and \cite{pla16b} with the same profile validates our procedure.\\

The distributions shown in Fig. \ref{fig:results_no_prior} correspond to the main results of the analysis concerning the impact of a modification of the mean normalized pressure profile on the estimation of cosmological parameters from the analysis of the tSZ power spectrum. We note that the distributions of the $F$-parameter associated with the $P_h$ and $P_l$ profiles are centered on values that are respectively significantly lower and higher than the average distribution obtained using the $P_m$ profile. This shows the high dependence of $F$ with the mean normalized pressure profile considered in the cosmological analysis.\\

The cosmological parameters estimated through the analysis that combines BAO measurements\footnote{See for example \citep{and14} for more details on the BAO analysis.} with the observation of the CMB primary anisotropies do not depend on the value of the hydrostatic bias parameter. Therefore, we compute the corresponding distribution of $F$ from the Markov chains of the \planck\ CMB analysis\footnote{Downloaded from the \planck\ legacy archive: \url{https://pla.esac.esa.int/\#cosmology}} by fixing the value of the hydrostatic bias parameter to the mean of its uniform prior distribution that is $b=0.2$. The distribution that we obtain is represented in purple in Fig. \ref{fig:results_no_prior}. It is enclosed between the distributions estimated using the profiles $P_m$ and $P_l$. This shows that a small deviation from the self-similar hypothesis, favouring a mean pressure profile of the cluster population slightly lower than the one constrained at low redshift, would resolve the tension between the cosmological constraints obtained from the analysis of the CMB primary anisotropies and the cluster abundance. This result shows that it is fundamental to study the properties of the pressure profile of galaxy clusters in different regions of the mass-redshift plane in order to identify a potential variation of the mean pressure profile. It is interesting to note that the results obtained by \cite{mcd14} from the analysis of 80 clusters selected from the SPT catalog and observed by \chandra\ indicate that the mean normalized pressure profile of galaxy clusters for redshift $0.6 < z < 1.2$ is slightly lower than the profile estimated at low redshift. The NIKA2 tSZ large program will allow us to see if this trend is also verified through observations of the tSZ effect in a similar redshift range.\\

The modification of the pressure profile used in our analysis results in a shift of the maximum likelihood of $F$ that is similar to the one caused by a variation of the hydrostatic bias \citep[\emph{e.g.}][]{pla16c}. This is expected because the overall amplitude of the mean normalized pressure profile, $P_0$, plays a role that is symmetric to the $P_{500}$ term in Eq. (\ref{eq:ClSZ_cosmo_analysis}) which depends on the hydrostatic bias $b$. This result shows that it is possible that the observed tension between the constraints of $\sigma_8$ and $\Omega_m$ resulting from the CMB analysis and from the analysis of the tSZ power spectrum could be partly explained by a modification of the mean pressure profile without requiring extreme values of the hydrostatic bias parameter.\\

Although the combined parameter $F$ is the only one that can be constrained by an analysis based solely on the study of the tSZ power spectrum, it is important to interpret the results obtained on the values of the $\sigma_8$ and $\Omega_m$ parameters. They are indeed fundamental quantities to define the model of the formation of large-scale structures in the universe. The three MCMC analyses described in Sect. \ref{sec:MCMC_planck} have also been realized by considering informative priors on the values of $\Omega_m$ and $b$ in order to obtain tighter constraints on the $\sigma_8$ and $\Omega_m$ parameters from the fit of the \planck\ tSZ power spectrum. We consider a Gaussian prior centered on 0.2 with a standard deviation of 0.08 for the hydrostatic bias given the averaged value of the current measurements of this parameter presented in \cite{sal18}. The Gaussian prior associated with the $\Omega_m$ parameter is defined to be a conservative constraint given by the BAO measurements \citep[see \emph{e.g.}][]{eis05} with a mean value of 0.3 and a standard deviation of 0.05. We represent the contours of the $\sigma_8$ and $\Omega_m$ parameters obtained at the end of these analyses in the left panel of Fig. \ref{fig:results_prior}. We note that the contours obtained by the \planck\ collaboration from the analysis of the tSZ power spectrum for a hydrostatic bias $b=0.2$ are located in the band constrained by our analysis based on the same profile $P_m$ (in grey). The thickness of the contours obtained at the end of our analysis is much larger than the one associated with the \planck\ contours because of the sampling performed on the $b$ parameter constrained by a Gaussian prior. We note that the difference between the means of the contours obtained with the $P_m$ and $P_l$ profiles is of the same order of magnitude as the $1\sigma$ width of each contour caused primarily by the uncertainty on the hydrostatic bias parameter. The current uncertainty on the hydrostatic bias as well as the potential variation of the mean normalized pressure profile of the cluster population compared to the low redshift profile therefore generates a significant systematic bias on the constraints of the $\sigma_8$ and $\Omega_m$ parameters obtained by the analysis of the tSZ power spectrum.\\ 

We also realize the same analysis considering a possible future hydrostatic bias prior of $b = 0.20 \pm 0.01$ as it was also done in the cluster count analysis detailed in \cite{pla16c}. The results are presented in the right panel of Fig. \ref{fig:results_prior} for the $P_m$ profile (grey) and for the same profile scaled down by 15\% (green). The cosmological constraints obtained with the $P_m$ profile present a 2-$\sigma$ discrepancy with the ones obtained from the analysis of the CMB primary anisotropies (purple). However, we show that a 15\% decrease of the amplitude of the mean normalized pressure profile compared to the one measured at high mass and low redshift would reconcile the maximum likelihood values of $\sigma_8$ and $\Omega_m$ with a hydrostatic bias that is fully compatible with the current estimates.\\

Both the uncertainties on the hydrostatic bias parameter and the mean normalized pressure profile induce systematic effects that are much greater than the magnitude of the tension observed with the estimates of $\sigma_8$ and $\Omega_m$ from low and high redshift probes. It is therefore essential to precisely constrain the potential mass and redshift evolution of the ICM thermodynamic properties. If such a deviation from the self-similar formation hypothesis is identified, it will be necessary to include the mass and redshift dependence of $b$ and the $\mathds{P}$ profile in cosmological analyses in order to obtain precise constraints of the $\sigma_8$ and $\Omega_m$ parameters from the analysis of the tSZ power spectrum.

\section{Conclusions}\label{sec:Conclusions}

The analysis developed in this paper has shown that the mean normalized pressure profile considered in cosmological analyses based on the study of the power spectrum of the tSZ effect plays a major role in the final estimates of the cosmological parameters $\sigma_8$ and $\Omega_m$. The true mean normalized pressure profile of the whole cluster population may be different from the ones currently used in cosmological analyses if the cluster samples considered so far to estimate it are not representative of the cluster population. For example, this difference may be due to a mass and redshift evolution of ICM properties or selection biases in either or both the calibration and cosmological samples.\\
We have defined two extreme cases of mean normalized pressure profiles based on the current knowledge of the pressure and gas mass fraction profile distributions at low redshift. We have used these profiles in a MCMC analysis in order to derive cosmological constraints in the hypothetical scenario in which the true mean of the pressure profile distribution in the whole mass-redshift plane is respectively lower and higher than the one observed at high mass and low redshift.\\
We further consider a mean normalized pressure profile that is compatible with the distribution of the profiles observed at low redshift by \planck\ but with a 15\% decrease of its amplitude compared to the ones chosen in current cosmological analyses. This profile allows us to alleviate the tension observed between the estimates of cosmological parameters from the primary CMB anisotropies and cluster analyses without requiring extreme value of the hydrostatic bias parameter.\\
It is therefore essential to accurately characterize the mean normalized pressure profile of the population of galaxy clusters by exploring regions of the mass-redshift plane that are still poorly known in order to identify a potential deviation from the self-similar formation hypothesis.\\
The NIKA2 tSZ large program will provide valuable insights concerning the evolution of the parameters of the mean normalized pressure profile with redshift. It will also allow us to study the corresponding evolution of the slope and intrinsic scatter of the scaling relation. The latter could indeed have a major impact on the definition of the selection function in cluster count analyses and induce significant systematic effects on the estimates of cosmological parameters. Taking these systematic effects into account in cosmological analyses will eventually make it possible to validate or invalidate the disagreement currently observed between the constraints of $\sigma_8$ and $\Omega_m$ resulting from the analysis of CMB primary anisotropies and from the statistical properties of galaxy clusters.

%###############################################################################################
%##########################                       ACKNOWLEDGEMENTS                        ##########################%###############################################################################################
\section*{Acknowledgements}
We would like to thank the anonymous referee for the helpful comments and suggestions. This work has been partially funded by the ANR under the contracts ANR-15-CE31-0017. Support for this work was provided by NASA through SAO Award Number SV2-82023 issued by the Chandra X-Ray Observatory Center, which is operated by the Smithsonian Astrophysical Observatory for and on behalf of NASA under contract NAS8-03060. FR would like to thank M. Arnaud, J-B. Melin, and especially B. Bolliet for very useful and interesting discussions. We acknowledge funding from the ENIGMASS French LabEx.

\bibliographystyle{mnras}

\appendix

\section{Additional figures}\label{sec:A1}

\begin{figure*}
\centering
\includegraphics[height=14cm]{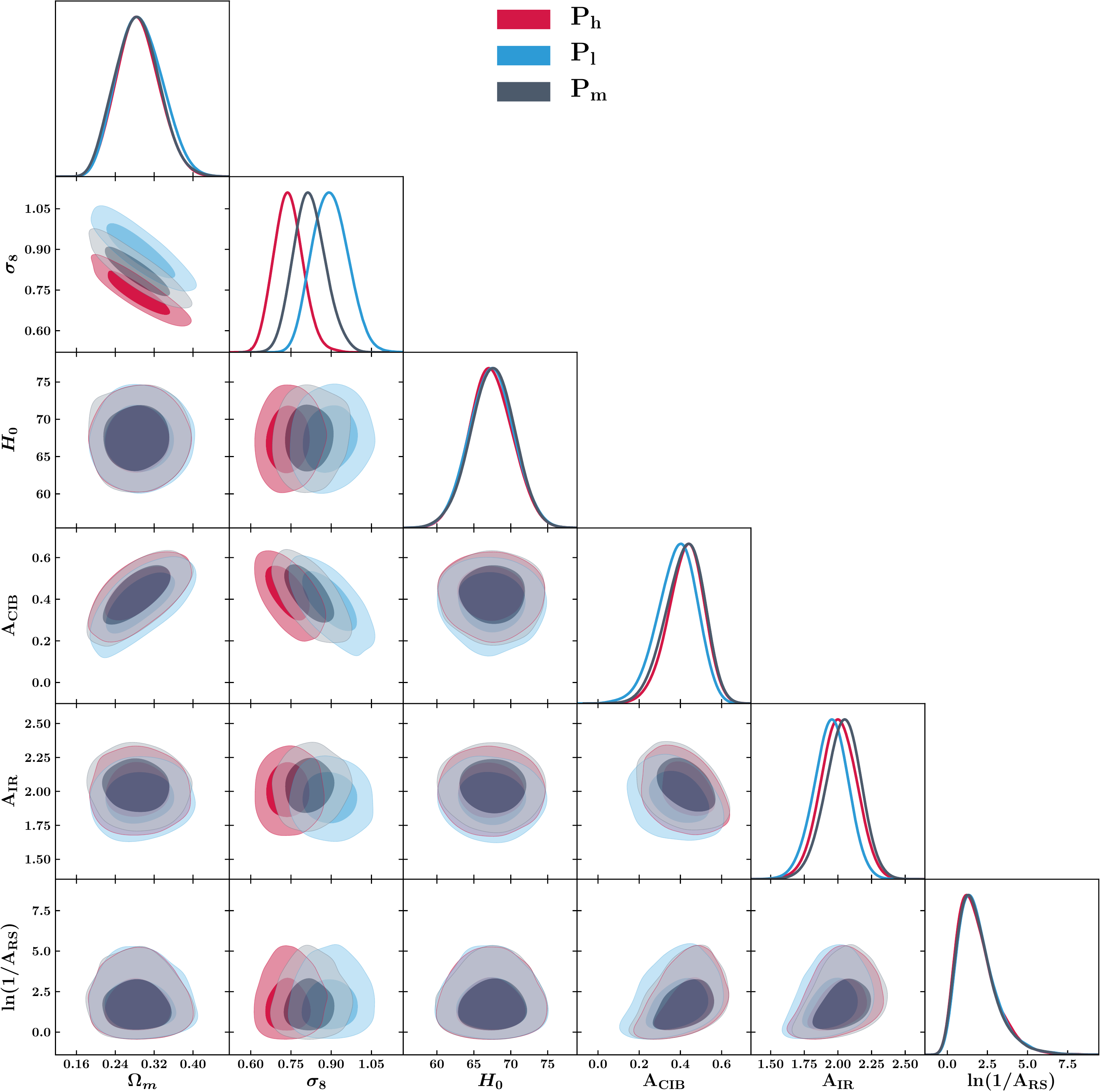}
\caption{Same as Fig. \ref{fig:contours_MCMC} comparing the results obtained with the three mean normalized pressure profiles defined in Sect. \ref{subsec:selection_P_prof}. We use the same color code as in Fig. \ref{fig:P_profiles_gas_frac} to show the constraints obtained with the $P_h$,  $P_l$, and $P_m$ profiles. We use a conservative Gaussian prior on the Universe matter density given by $\Omega_m = 0.30 \pm 0.05$.}
\label{fig:A1_compare}
\end{figure*}

\begin{figure*}
\centering
\includegraphics[height=6cm]{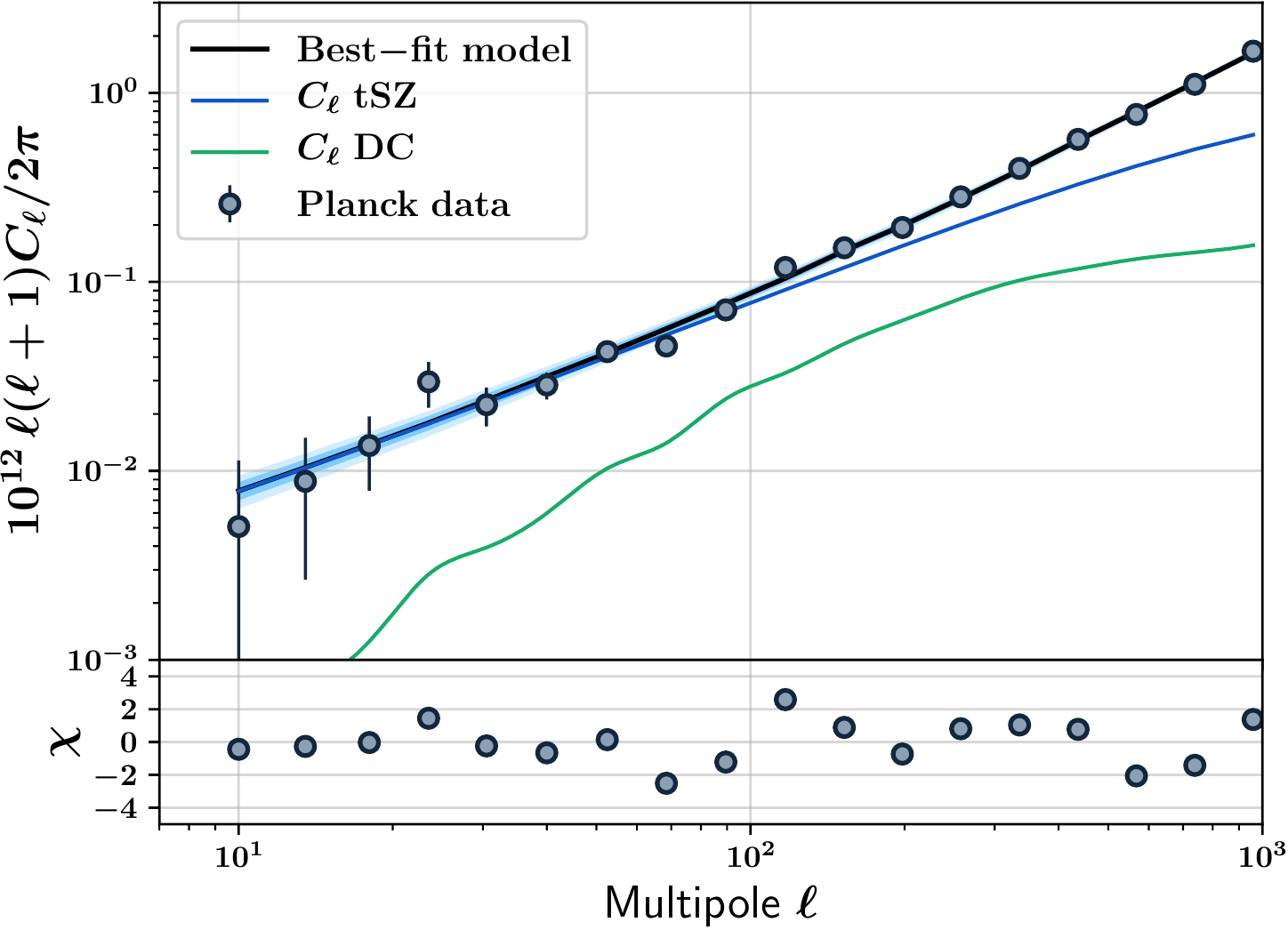}
\hspace{0.4cm}
\includegraphics[height=6cm]{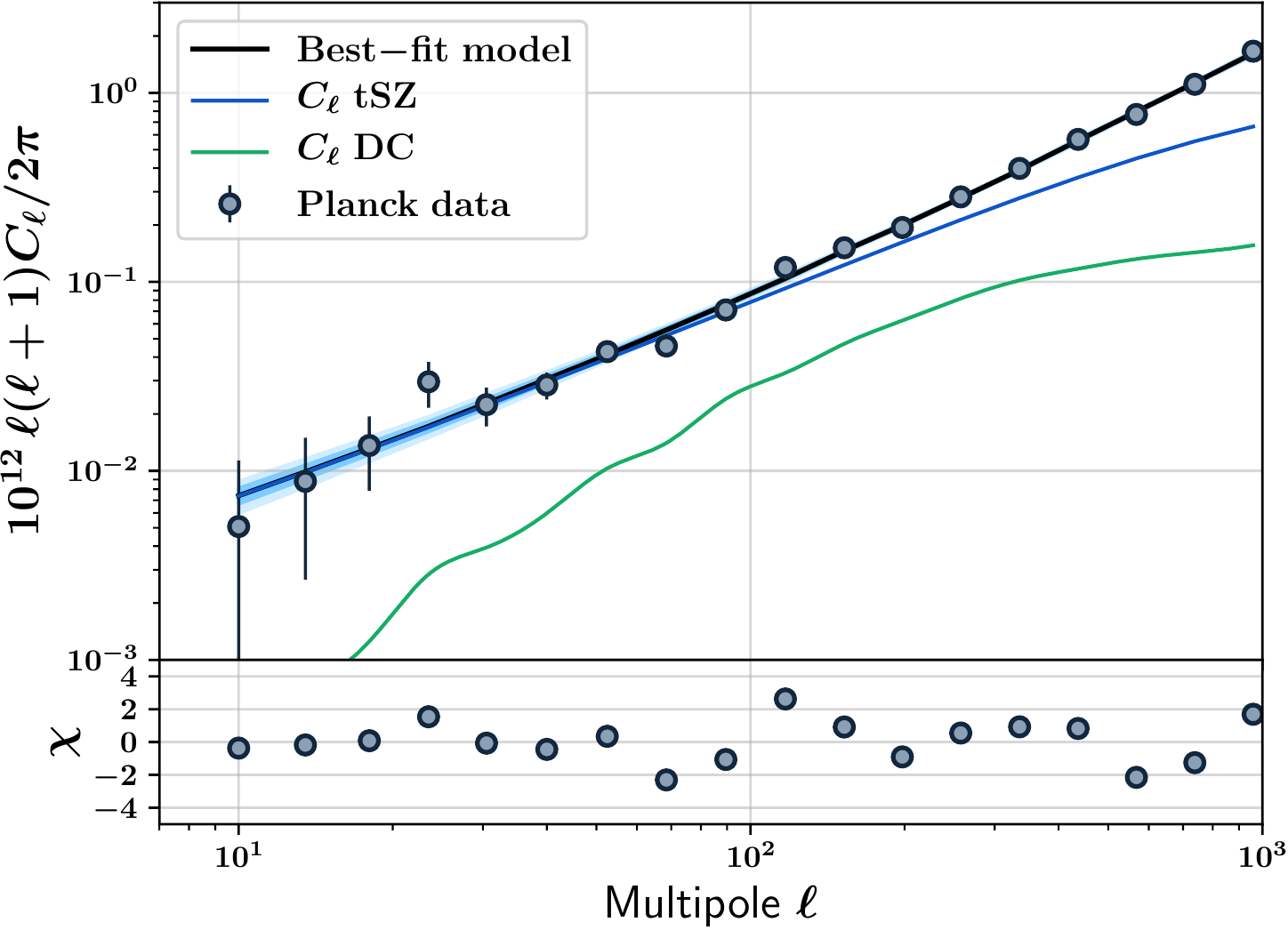}
\caption{Same as Fig. \ref{fig:best_fit_model} using the $P_l$ (left) and $P_h$ (right) profiles in the MCMC analyis of the \planck\ tSZ power spectrum.}
\label{fig:A2_bestfits}
\end{figure*}

\label{lastpage}
\end{document}